\documentclass[a4paper,12pt]{article}
\usepackage[colorlinks,pdfpagelabels,pdfstartview=FitH,bookmarksopen=true,bookmarksnumbered=true,linkcolor=blue,plainpages=false,hypertexnames=false,citecolor=blue, urlcolor=blue]{hyperref}

\usepackage{amsfonts}
\usepackage{amsmath}
\usepackage{graphicx}
\usepackage{amssymb}
\usepackage{longtable}
\usepackage{authblk}
\usepackage{pdflscape}
\usepackage{rotating}
\usepackage[round]{natbib}
\usepackage{booktabs}
\usepackage{siunitx}

\usepackage[hmargin=2.5cm,vmargin=3.5cm]{geometry}

\usepackage{setspace}
\usepackage{chngcntr}
\usepackage{verbatim}
\usepackage{multirow}
\usepackage{subcaption}
\usepackage{chngcntr}

\usepackage[title]{appendix}
\usepackage[gen]{eurosym}
\usepackage{comment}
\usepackage[bottom]{footmisc}
\usepackage{gensymb}
\usepackage{textcomp,mathcomp}
\usepackage{wrapfig}
\usepackage{pdflscape}
\usepackage{rotating}
\usepackage{epstopdf}
\usepackage{graphicx}
\usepackage{caption}
\usepackage{subcaption}
\usepackage{longtable,tabularx,ltxtable,ragged2e}
\usepackage{titlesec}
\titleformat{\section}
  {\normalfont\large\bfseries}{\thesection}{1em}{}
\titleformat{\subsection}
  {\normalfont\normalsize\bfseries}{\thesubsection}{1em}{}
\titleformat{\subsubsection}
  {\normalfont\normalsize\bfseries}{\thesubsubsection}{1em}{}

\begin{document}

\pagenumbering{arabic}
\author[1] {Haroon Mumtaz}
\affil[1]{Queen Mary University of London}

\author[2] {Sofia Velasco}
\affil[2]{Banco de España}

\title{ A Dynamic Factor Model for Level and Volatility
\thanks{\footnotesize Corresponding author: Haroon Mumtaz (\href{mailto:h.mumtaz@qmul.ac.uk}{h.mumtaz@qmul.ac.uk}). The views expressed in this paper are those of the authors and do not represent those of Banco de Espa\~{n}a or the ESCB.}
}

\date{This version: \today.}
\maketitle

\begin{abstract}
\setlength{\parskip}{2pt}
This paper develops a dynamic factor model in which common level and volatility factors evolve jointly, allowing conditional means and variances to interact endogenously within a large-information setting. The joint evolution of these factors provides a tractable framework for modeling risk, as fluctuations in volatility affect both the dispersion and the location of outcomes, generating state-dependent and asymmetric tail risks in predictive distributions. Volatility is captured by latent common factors that drive co-movement in second moments across a large panel, while heavy-tailed idiosyncratic shocks absorb transitory outliers and isolate persistent uncertainty dynamics. The framework embeds these interactions directly within a factor structure, allowing risk to arise endogenously from the joint dynamics of the system rather than being imposed through reduced-form approaches. Empirically, the model delivers systematic improvements in density forecast accuracy, particularly in the tails of the predictive distribution and at medium horizons. An application to international inflation highlights a dominant global level component in advanced economies and stronger regional and volatility contributions in emerging and developing economies, pointing to substantial heterogeneity in the role of uncertainty across countries.
\end{abstract}

\vspace{-1ex}

\clearpage

\section{\label{intro}Introduction}

Macroeconomic uncertainty is widely viewed as a key driver of business-cycle fluctuations, financial conditions, and downside risks to real activity. A central lesson from recent research is that its macroeconomic relevance depends not only on time variation in conditional volatility, but also on whether fluctuations in volatility are systematically related to the conditional mean. When conditional means and variances co-move, shocks to uncertainty affect not only the dispersion of future outcomes but also their location, generating state-dependent and asymmetric tail risks in predictive distributions. This mechanism is closely related to volatility-in-mean and volatility-feedback effects studied in macro-finance models (e.g., \citealp{MUMTAZ201810,shin2020new,CaldaraSVOL}) and gives rise to Growth-at-Risk--type dynamics, whereby downside risks increase disproportionately in high-volatility states.

A growing empirical literature measures uncertainty through time variation in the unpredictable component of macroeconomic variables and documents that conditional variances often contain a systematic common component across series (e.g., \citealp{Carriero2018,mumtaz2021evolving,Castelnuovo2025}). When volatility rises simultaneously across many variables, such movements are unlikely to be purely idiosyncratic and instead reflect an aggregate uncertainty component. In parallel, the financial econometrics literature models large time-varying covariance matrices using low-dimensional latent volatility factors (e.g., \citealp{aguilar2000bayesian,han2006asset,lopes2007factor}), providing a parsimonious representation of heterogeneous co-movement in second moments while avoiding over-parameterization.

A key modeling question concerns how to represent volatility in large systems. While some approaches impose a single common volatility process, such specifications can be overly restrictive because they imply uniform shifts in dispersion across heterogeneous variables. More flexible structures allow for richer covariance dynamics while remaining computationally tractable. In this context, \citet{chan2023comparing} shows that, in large systems, factor stochastic volatility specifications provide a superior fit to predictive densities relative to models based on a single common volatility component. Complementary evidence in \citet{Castelnuovo2025} emphasizes that volatility dynamics are themselves heterogeneous, reflecting a combination of aggregate, sectoral, and idiosyncratic components. Taken together, these findings support the use of factor-based volatility structures as a flexible and empirically relevant way to model uncertainty in data-rich environments.

Despite these advances, most large-information macroeconomic models treat volatility either as series-specific or as exogenous to the evolution of macroeconomic activity. As a result, they capture time variation in uncertainty but do not allow it to interact endogenously with the forces driving the conditional mean. By contrast, a growing VAR-based literature shows that such interactions are empirically important for state-dependent dynamics and macroeconomic risk. In particular, \citet{MUMTAZ201810} and \citet{CaldaraSVOL} show that when volatility co-moves with economic activity, shocks to uncertainty affect not only dispersion but also expected outcomes, shaping the entire predictive distribution rather than only its variance. Recent work on macroeconomic risk likewise emphasizes the importance of modeling tail behavior directly in the predictive distribution \citep{caldara2024risk}.

This paper brings these strands of the literature together in a unified large-information framework. We develop a dynamic factor model in which common level and volatility factors evolve jointly within a VAR system. Volatility is captured by a small number of latent common factors that summarize co-movement in second moments across a large panel of macroeconomic variables, while the joint transition equation allows level and volatility dynamics to interact endogenously. This structure embeds volatility-in-mean effects directly within a large-information factor model, allowing uncertainty shocks to affect both conditional variances and expected macroeconomic outcomes through the joint dynamics of the latent factors.

A key feature of the model is that level and volatility factors follow a joint VAR process, allowing innovations to economic activity and volatility to be contemporaneously correlated. In particular, periods of declining economic activity can be associated with increases in volatility. When such correlations are present, the predictive distribution shifts in both location and dispersion, generating asymmetric tail behavior without requiring large increases in overall volatility. When the correlation between level and volatility shocks is negative, downturns are accompanied by increases in uncertainty, so that the left tail of the predictive distribution moves more strongly than the right tail, leading to a disproportionate increase in downside risk. The model therefore provides a unified and data-rich framework for capturing Growth-at-Risk--type dynamics.

We evaluate the empirical relevance of this structure in two applications. First, using U.S.\ macroeconomic data, we compare the forecasting performance of the level--volatility DFM to that of a benchmark dynamic factor model that allows for stochastic volatility at the series level but excludes common volatility factors. Incorporating common volatility factors yields systematic and statistically significant improvements in density forecast accuracy, concentrated in the tails of the predictive distribution and at medium horizons. These gains are most pronounced for employment, inflation, and the federal funds rate, and remain robust when the COVID-19 period is excluded. Effects on point forecasts are negligible, indicating that the model's primary contribution lies in capturing the evolution of macroeconomic risk rather than the conditional mean.

Second, we apply the framework to a decomposition of global and regional inflation dynamics. The joint level--volatility structure reveals a dominant global level component in advanced economies, while emerging market and developing economies exhibit stronger regional effects and more pronounced volatility contributions. Cross-country heterogeneity is particularly pronounced along the volatility dimension, a feature that standard mean-based decompositions are not designed to capture.

Overall, the contribution of the paper is to provide a tractable large-information framework in which volatility is modeled through latent factors that allow for heterogeneous co-movement across series, while level and volatility dynamics interact endogenously to generate state-dependent and asymmetric tail risks in predictive distributions. By combining insights from the factor stochastic volatility literature, the macroeconomic uncertainty literature, and the volatility-in-mean literature, the proposed approach delivers a unified framework for analyzing and forecasting macroeconomic tail risks.

The remainder of the paper is organized as follows. Section~\ref{empirical} introduces the level--volatility DFM framework, discusses estimation, and reports a Monte Carlo simulation on finite-sample recovery. Section~3 presents the U.S.\ forecasting application, including the estimation of common uncertainty factors and the evaluation of density and tail-risk forecasts. Section~\ref{Decomp} studies global and regional inflation dynamics. Section~\ref{conclu} concludes. Additional technical details and supplementary results are reported in the Appendix.

\section{\label{empirical} A factor model with joint level and volatility dynamics}
In this section, we present the econometric model used to extract common level and volatility factors and describe the estimation procedure. We also evaluate the model’s ability to recover joint level and volatility dynamics in finite samples through a Monte Carlo simulation exercise calibrated to the proposed state-space system. The simulation assesses whether the estimator accurately captures persistent, interacting mean and volatility dynamics without imposing restrictive distributional assumptions. Appendix~\ref{app:model} provides additional technical details.
\subsection{Model specification}
\noindent
To jointly estimate common level and volatility factors, we employ a factor model with time-varying volatility. The observation equation is given by
\begin{equation}
X_{it} = B_i f_t + v_{it}
\label{eq6}
\end{equation}
where $\underset{N \times 1}{X_t}$ collects the $i = 1,\dots,N$ observable variables. As in the canonical DFM of \cite{stock-watson-02} or the FAVAR framework of \cite{bernanke-boivin-eliasz-05}, observable variables are driven by a lower-dimensional set of common level factors $\underset{J \times 1}{f_t}$ to which they are associated via the loading matrix $\underset{N \times J}{B}$.

Departing from the standard specification, we allow for serial correlation and heteroskedasticity in the idiosyncratic component $\underset{N \times 1}{v_t}$. Each element of $v_t$ follows an AR($L^{v}$) model with heteroskedastic shocks $\epsilon_{it}$:
\begin{equation}
v_{it} = \sum_{l=1}^{L^{v}} \rho_{i,l} v_{i,t-l} + \epsilon_{it}, \hspace{2mm} \epsilon_{it} \sim N(0, r_{it})
\label{eq51}
\end{equation}
where the time-varying variance $r_{it}$ governs the unpredictable component of $X_{it}$ and can be interpreted as a series-specific measure of uncertainty. The link between unpredictability and uncertainty underlies a large empirical literature (e.g., \citealp{jurado2015measuring,carriero2016common,jo2019macroeconomic}). Building on this notion of uncertainty, a number of studies examine its cross-sectional dimension and document that conditional variances contain a systematic common component across endogenous variables (see, e.g., \citealp{Carriero2018,mumtaz2021evolving,Castelnuovo2025}). The intuition is that if volatility rises simultaneously across many series, such movements are unlikely to be purely idiosyncratic and instead reflect an aggregate uncertainty component.

Moreover, a strand of the financial econometrics literature models time variation in large covariance matrices through a small number of latent volatility factors (e.g., \citealp{aguilar2000bayesian,han2006asset,lopes2007factor}). In these frameworks, the time-varying covariance matrix is parameterized via a low-dimensional factor structure, so that both variances and covariances inherit their dynamics from a small set of common stochastic volatility processes. This strategy avoids over-parameterizing the covariance structure. The resulting factor decomposition isolates common and idiosyncratic volatility components, attributing systematic co-movement in conditional variances to shared volatility drivers. This perspective supports the view that a latent volatility factor may be shared across endogenous variables and convey independent information about aggregate uncertainty.

Motivated by these modeling strategies, we allow the variance of the idiosyncratic shocks to depend on a set of common volatility factors $\underset{k\times1}{\mathcal{F}_{t}}$ with loadings $\underset{N \times k}{\mathcal{B}}$:

\begin{equation}
r_{it} = \frac{e^{\mathcal{B}_{i}\mathcal{F}_{t}}}{\lambda_{it}},
\label{eq4}
\end{equation}

The term $e^{\mathcal{B}_{i}\mathcal{F}_{t}}$ captures persistent and systematic movements in volatility driven by common factors, while $\lambda_{it}$ governs transitory, series-specific scale variation. To accommodate occasional extreme realizations, idiosyncratic innovations are modeled using a scale-mixture-of-normals representation \citep{Geweke1993}. Conditional on the latent scale variable, the disturbances are Gaussian; marginally, they follow a Student-$t$ distribution (see also \citealp{CHIU20171124}). This specification isolates infrequent large shocks from persistent volatility dynamics and prevents them from being misinterpreted as aggregate uncertainty movements.

While this specification for $r_{it}$ is closely related to \cite{Carriero2018} and \cite{Castelnuovo2025}, our framework differs in the way levels and volatilities interact. We treat the level factors $f_t$ and the volatility factors $\mathcal{F}_t$ as jointly endogenous state variables and allow for contemporaneous and lagged interactions between them.

Specifically, we postulate the following joint transition equation for the stacked factor vector $F_{t}=\begin{pmatrix} f_{t} \\ \mathcal{F}_{t} \end{pmatrix}$, which evolves according to a VAR($L^F$) process:
\begin{eqnarray}
F_{t} &=& c+\sum_{j=1}^{L^F}\beta_{j}F_{t-j}+e_{t},  \hspace{2mm} e_{t} \sim N(0,\Omega). \label{eq1}
\end{eqnarray}

The key modeling feature is that the covariance matrix $\Omega$ allows for non-zero off-diagonal elements linking innovations to level and volatility factors. This implies that shocks to volatility can directly affect the evolution of the level factors through the VAR dynamics. In contrast to standard stochastic volatility specifications, where volatility evolves independently of the conditional mean, the present framework embeds a volatility-in-mean mechanism within a factor model. This interaction is central for generating asymmetric tail risks in predictive distributions and time variation in macroeconomic tail risks. In this sense, the framework embeds volatility-in-mean dynamics in a large-information DFM setting, in the spirit of \cite{MUMTAZ201810} and \cite{CaldaraSVOL}, while permitting dynamic correlation between conditional means and conditional variances.
\subsection{Estimation}

\subsubsection{Priors and starting values}

To define priors for the VAR coefficients, we follow \citet{banbura2010large} and implement the dummy observation approach to impose a Minnesota-type prior \citep{Doan1983,SimsZha1998}. The overall shrinkage is controlled by the tightness parameter, which we set to $\tau=0.1$, a standard calibration for U.S.\ data \citep{alessandri2017financial}. The covariance matrix $\Omega$ is factorized as $\Omega=A^{-1}HA^{-1'}$, with Gaussian priors on the non-zero elements of $A$ and inverse-Gamma priors on the diagonal elements of $H$.

Factor loadings $B_i$ and $\mathcal{B}_i$ receive Gaussian priors centered on preliminary PCA estimates. For the scale parameters governing heavy-tailed idiosyncratic shocks, we assume $\lambda_{it}\sim\Gamma(1,\nu_i)$ following \citet{Geweke1993}, implying a scale-mixture representation and conditionally Gaussian Student-$t$ innovations. The degrees-of-freedom parameter $\nu_i$ is treated hierarchically, and persistence parameters $\rho_i$ are assigned Gaussian priors.

\subsection{Posterior simulation}

We approximate the joint posterior distribution of the parameters and latent states $(F_t,\Gamma,\Omega,B_i,\mathcal{B}_i,\rho_i,\lambda_{it},\nu_i)$ using a Gibbs sampling algorithm. Full derivations are provided in Appendix~\ref{app:gibbs}. At each iteration, conditional on the current draws of the remaining parameters and states (denoted generically by $\Psi$), the sampler cycles through the following blocks:

\begin{enumerate}

\item \textbf{VAR coefficients and covariance matrix $F(\Gamma,\Omega|\Psi)$.}
Given the factor draws $F_t$, the transition equation \eqref{eq1} is a standard VAR. Under the priors described above, the conditional posterior for the coefficients $\Gamma$ is Gaussian. The covariance matrix $\Omega$ is sampled via its decomposition $\Omega=A^{-1}HA^{-1'}$, with Gaussian posteriors for the non-zero elements of $A$ and inverse-Gamma posteriors for the diagonal elements of $H$. For computational efficiency, estimation can be implemented equation-by-equation as in \citet{CARRIERO2022506}.\footnote{In the implementation we rely on the equation-by-equation algorithm. Under conjugate priors, this approach is numerically equivalent to the standard joint Normal–inverse-Wishart sampling scheme, differing only in computational efficiency.}

\item \textbf{Factor loadings $F(B_i|\Psi)$.}
Conditional on $(f_t,\rho_i,r_{it})$, the measurement equation can be written as a linear regression after GLS transformation. The conditional posterior of $B_i$ is Gaussian with standard closed-form expressions for mean and variance (see, e.g., \citealp{kim-nelson-98}).

\item \textbf{Idiosyncratic persistence $F(\rho_i|\Psi)$.}
Given $v_{it}$ and $r_{it}$, the AR($L^v$) specification in \eqref{eq51} reduces to a linear regression with heteroskedastic errors. After GLS transformation, the conditional posterior for $\rho_i$ is Gaussian.

\item \textbf{Volatility loadings $F(\mathcal{B}_i|\Psi)$.}
Conditional on $(\mathcal{F}_t,\lambda_{it})$, volatility is nonlinear in $\mathcal{B}_i$. We therefore employ a random-walk Metropolis--Hastings step. Candidate draws are evaluated using the Gaussian likelihood implied by the transformed measurement equation.

\item \textbf{Scale parameters $F(\lambda_{it}|\Psi)$.}
Given $\epsilon_{it}$ and $\tilde r_{it}=e^{\mathcal{B}_i\mathcal{F}_t}$, the conditional posterior of $\lambda_{it}$ is Gamma \citep{koop03}, reflecting the scale-mixture representation of Student-$t$ disturbances.

\item \textbf{Degrees of freedom $F(\nu_i|\Psi)$.}
The conditional posterior of $\nu_i$ is non-standard. We draw $\nu_i$ using a Metropolis--Hastings step based on its Gamma prior and the likelihood implied by the scale-mixture representation.

\item \textbf{Latent factors $F(F_t|\Psi)$.}
Conditional on all parameters, the model admits a state-space representation with nonlinear volatility. We draw the stacked factor vector $F_t=(f_t',\mathcal{F}_t')'$ using a particle Gibbs sampler with ancestor sampling as in \citet{JMLR:v15:lindsten14a}, which accommodates multiple lags in the transition equation.

\end{enumerate}

\subsection{Monte Carlo Simulation}

To assess the ability of the proposed framework to recover joint level and volatility dynamics, we conduct an overall Monte Carlo simulation based on data generated from the state-space system in equations \eqref{eq6}--\eqref{eq1}. The data-generating process (DGP) is calibrated to replicate key features of the empirical specification, including persistent level and volatility factors that evolve jointly and allow for correlation between innovations to the conditional mean and conditional variance, consistent with a volatility-in-mean mechanism.

The simulation evaluates whether the estimation procedure can recover the key features of the DGP. In particular, we focus on (i) the latent level and volatility factors, (ii) the persistence of each component, and (iii) the joint dynamics linking innovations to levels and volatility.

We present results based on a single realization of the DGP. The findings indicate that the model successfully recovers the common volatility factor and its interaction with the level factors. In particular, the estimated system reproduces the asymmetric features of the predictive distribution implied by the DGP, confirming that the methodology is able to capture joint location--scale dynamics rather than attributing them to spurious idiosyncratic variation.

For brevity, the full set of simulation results and additional details on the simulation set-up are reported in Appendix~\ref{app:MCsim}.

\section{\label{FEV}Forecasting U.S. macroeconomic variables}
This section evaluates the pseudo out-of-sample forecasting performance of the proposed level--volatility DFM using U.S.\ data. In this first application, forecasts are benchmarked against those from a linear dynamic factor model (DFM), which serves as a standard reference specification in large-scale macroeconomic forecasting (see, e.g., \citealp{giannone2008nowcasting}). To ensure a fair comparison, the benchmark DFM features linear factor dynamics but allows for serial correlation and stochastic volatility in the idiosyncratic components of the measurement equation. However, the benchmark abstracts from common volatility factors and from interactions between conditional means and variances.

The estimation sample begins in 1965Q1 and spans more than six decades, covering several distinct macroeconomic regimes, including the Great Moderation, the dot-com boom and bust, the Global Financial Crisis, and the COVID-19 pandemic. The associated shifts in macroeconomic dynamics and uncertainty provide a demanding environment for forecast evaluation and a natural setting to assess the gains from jointly modeling levels and volatilities. The empirical design follows the large-dimensional forecasting literature, using an expanding-window pseudo out-of-sample scheme and a broad panel of U.S.\ macroeconomic and financial variables from the quarterly FRED-QD database of \citet{McCrackenNg2021}.

In addition to standard point and density forecast evaluation, we assess the model’s ability to capture tail risks following \citet{carriero2024capturing}, using tail-focused scoring rules and comparisons with quantile regression benchmarks. This exercise allows us to evaluate whether the model improves the characterization of downside and upside risks in the predictive distribution.

\subsection{Forecast design}

The forecasting exercise uses quarterly U.S.\ macroeconomic data from the FRED-QD database. FRED-QD contains 248 quarterly series. From this set, we select $N=105$ variables to replicate the information content of \citet{Stock-Watson-12} and to construct a broad information set spanning real activity, labor market conditions, prices, monetary aggregates, interest rates, and financial variables. A complete list of variables and the corresponding transformation codes is reported in Appendix~\ref{app:data}.

We focus on a set of core macroeconomic aggregates as target variables for forecast evaluation: real output (GDPC1), non-farm employment (PAYEMS), consumer prices (CPIAUCSL), and the effective federal funds rate (FEDFUNDS). Forecasts are generated using an expanding-window scheme. The estimation sample begins in 1965Q1 and is recursively updated each quarter over the hold-out period. We report results for short- and medium-horizon forecasts, corresponding to 1-year- and 2-year-ahead horizons.

All variables are transformed to stationarity using the transformation codes provided with FRED-QD. Density forecasts are constructed at each forecast origin by propagating posterior uncertainty through the predictive distribution.

Forecast performance of the proposed level–volatility DFM is evaluated relative to a benchmark DFM. In the benchmark specification, serial correlation and heteroskedasticity are allowed for at the level of the observed series through the idiosyncratic components of the measurement equation, whose dynamics are modeled explicitly using autoregressive processes with stochastic volatility. Conditional on the latent volatility states, the model remains Gaussian and linear in levels.

Factor dynamics in the benchmark DFM are governed by a linear VAR with homoskedastic innovations, matching the specification used for the level factors in the proposed model. As a result, the two frameworks differ primarily in their treatment of uncertainty: while the benchmark accommodates persistent and time-varying idiosyncratic volatility at the series level, it abstracts from common volatility factors and from interactions between conditional means and conditional variances. Further details of the benchmark specification are provided in Appendix~\ref{app:dfm}.

\subsection{Common Uncertainty}\label{commonunc}

\begin{figure}[!htbp]
    \centering
\includegraphics[width=\textwidth]{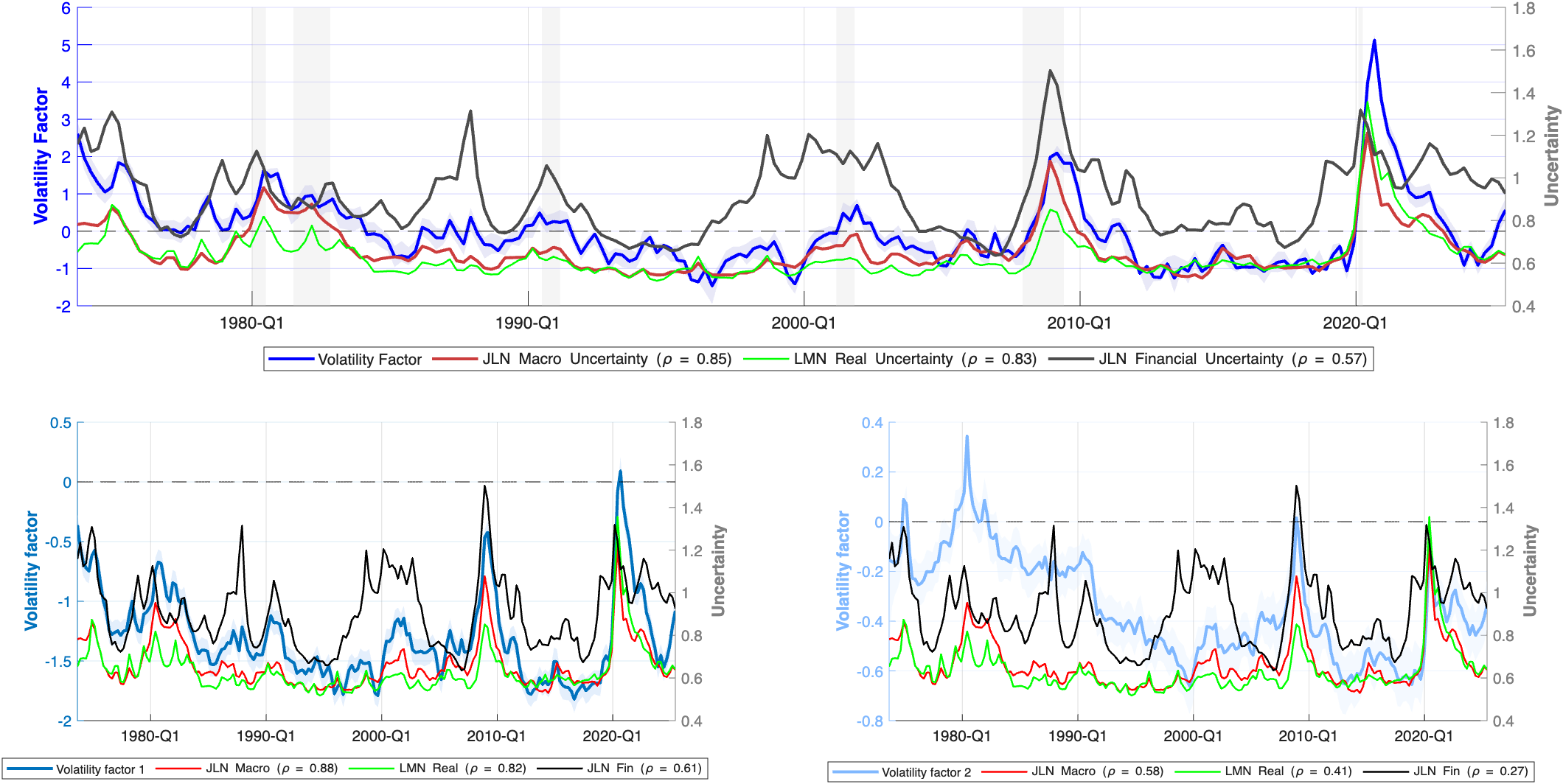}
        \caption{Common Volatility}
    \label{fig:uncertainity}
    \caption*{\footnotesize \textbf{Notes:} The figure reports estimates from the U.S. specification with two level factors and one volatility factor in the first row and two level and two volatility factors in the second row. The solid blue lines denotes the posterior median of the common volatility factor, and the shaded area denotes the 68\% credible intervals. The red and gray lines report the macroeconomic and financial uncertainty measures of \cite{jurado2015measuring} and \cite{ludvigson2021uncertainty}, normalized for comparability. Gray vertical bands indicate NBER recession periods.}
\end{figure}

Figure~\ref{fig:uncertainity} reports the estimated common volatility factors from the baseline specifications with two level factors and one volatility factor (first row) and two level factors and two volatility factors (second row), based on full-sample estimates. To assess their economic interpretation, we compare the estimated factors with widely used measures of macroeconomic uncertainty, following \citet{Castelnuovo2025}. Specifically, we consider the macroeconomic and financial uncertainty indexes of \citet{jurado2015measuring} and \citet{ludvigson2021uncertainty}.

In the single-volatility-factor specification (top panel), the estimated factor closely comoves with external uncertainty measures and displays pronounced spikes during major episodes of macroeconomic stress, most notably the Global Financial Crisis and the COVID-19 pandemic. The correlation with the macroeconomic and real uncertainty indexes is strong (exceeding $\rho=0.80$), while the association with financial uncertainty is more moderate ($\rho=0.57$). This pattern indicates that the factor captures a broad common component in the unpredictable variation of macroeconomic series rather than fluctuations specific to financial markets.

The two-volatility-factor specification (bottom panels) provides a more granular decomposition of uncertainty. The first volatility factor closely replicates the single-factor estimate, with a correlation of $0.97$, and remains strongly associated with macroeconomic uncertainty ($\rho=0.88$). This suggests that it represents the dominant aggregate uncertainty component. The substantial comovement between the two volatility factors ($\rho=0.67$) further indicates that they capture related dimensions of aggregate volatility.

In contrast, the second volatility factor exhibits weaker comovement with macroeconomic ($\rho=0.58$) and real uncertainty ($\rho=0.41$), and only a limited association with financial uncertainty ($\rho=0.27$). Its lower correlation with the single-factor estimate ($\rho=0.63$) points to a more secondary and less pervasive source of volatility.

Overall, the evidence indicates that the common volatility extracted from the FRED-QD panel primarily reflects macroeconomic uncertainty, consistent with \citet{Castelnuovo2025}. The dominant factor is stable across specifications and closely aligned with standard measures of macroeconomic uncertainty, while additional volatility factors capture more heterogeneous and less economically interpretable components of second-moment dynamics.

Building on the intuition developed in \cite{caldara2024risk}, we illustrate how the interaction between level and volatility dynamics generates asymmetric risks in the predictive distribution. In frameworks with volatility-in-mean effects, fluctuations in uncertainty can affect not only the dispersion of macroeconomic outcomes but also their conditional mean. In our model this mechanism arises naturally because the level and volatility factors evolve jointly within a VAR system. As a result, shocks to economic activity and shocks to volatility can propagate across the system and become contemporaneously correlated. When this correlation is negative, declines in the level factor tend to be accompanied by increases in volatility. In such episodes the predictive distribution shifts leftward while simultaneously widening. This joint location–scale adjustment produces asymmetric movements in the distribution’s tails, with the left tail expanding more strongly than the right tail. To illustrate this mechanism empirically, we examine the behavior of predictive densities for U.S.\ GDP growth around the Global Financial Crisis using estimates from the baseline model.

\begin{figure}[!htbp]
 \centering
 \includegraphics[width=\textwidth]{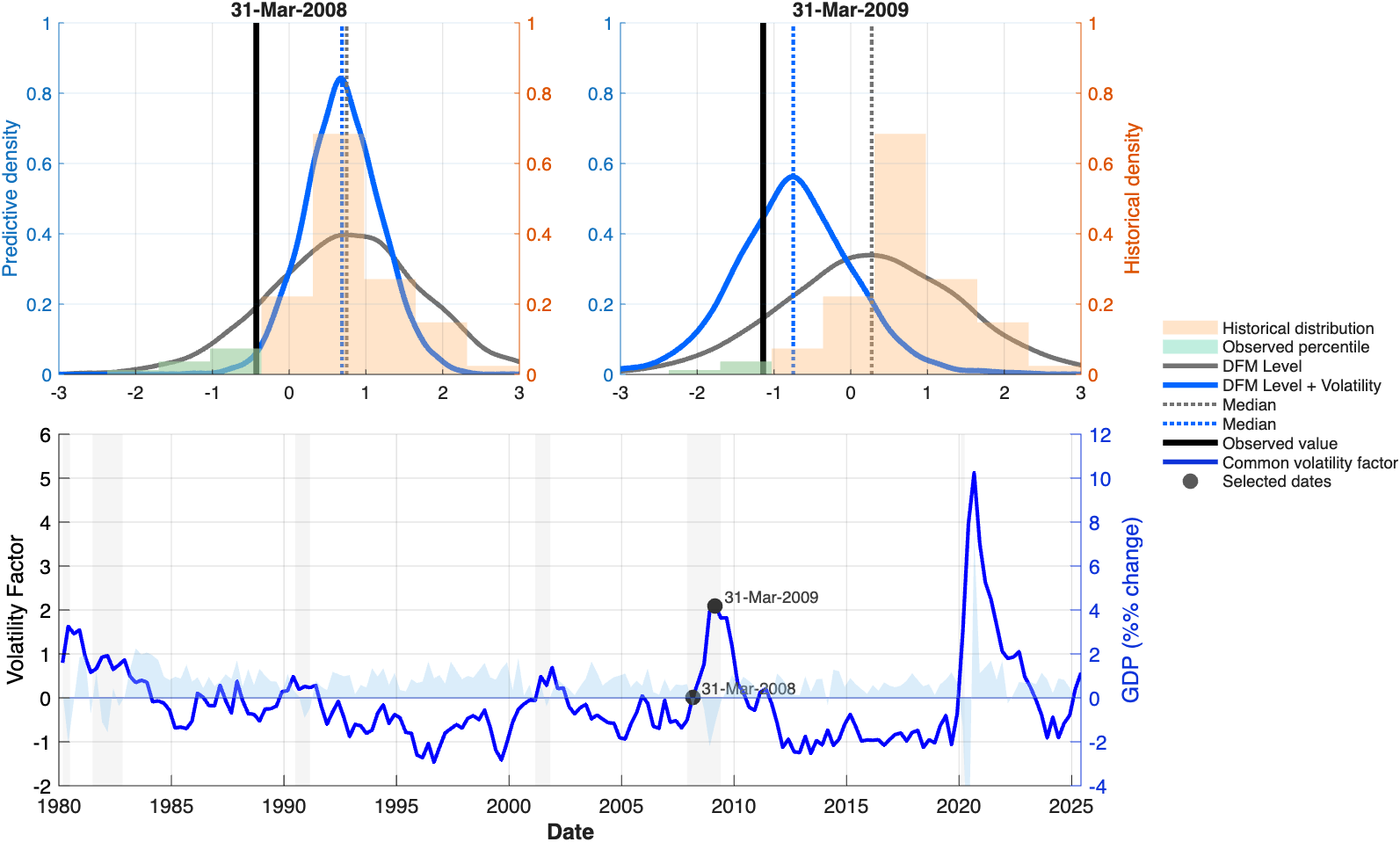}
 \caption{Output growth predictive densities and common volatility}
 \caption*{\footnotesize \textbf{Notes:} The upper panels report one-step-ahead predictive densities for GDP growth at selected dates around the Global Financial Crisis. The blue line denotes the predictive density from the level–volatility DFM, while the grey line corresponds to the benchmark DFM without a common volatility factor. Vertical dashed lines indicate the predictive medians and the solid black line denotes the realized outcome. The shaded histogram represents the historical distribution of GDP growth and the green bar highlights the percentile of the realized observation. The lower panel reports the estimated common volatility factor (LHS) together with the selected dates used in the density plots and the realized GDP over that period in light blue (RHS). }
 \label{fig:densities}
\end{figure}

Figure \ref{fig:densities} illustrates the mechanism through which the level–volatility DFM generates asymmetric risks. The lower panel reports the estimated common volatility factor together with realized GDP growth, while the upper panels display the corresponding one-step-ahead predictive densities at selected dates around the Global Financial Crisis. Because the level and volatility factors jointly evolve according to a VAR process, shocks to economic activity and uncertainty are allowed to interact contemporaneously.

In March 2008 the volatility factor is close to its historical average. Accordingly, the predictive medians from the benchmark DFM and the level–volatility DFM are nearly identical, as shown in the left panel. The realized GDP outcome, however, is unusually weak—around the 7th percentile of its historical distribution. This realization, together with similarly unusual realizations in other macroeconomic variables during the same period, feeds into the volatility factor through the joint VAR dynamics, generating a sharp increase in common volatility over the subsequent quarters, as shown in the lower panel.

By March 2009 the volatility factor has risen substantially. The right panel shows that the predictive density implied by the level–volatility model shifts to the left relative to the benchmark specification and places considerably more mass in the lower tail of the distribution. As a result, the model is able to accommodate the extremely weak GDP realization—around the 3rd percentile of its historical distribution—without requiring unrealistically large volatility shocks. The figure therefore illustrates how the interaction between level and volatility factors generates asymmetric downside risks in the predictive distribution.

\subsection{Competing specifications and evaluation}

We evaluate predictive performance along two dimensions. First, we compare the proposed model to standard dynamic factor models using point and density forecast metrics. Second, we assess the ability of the model to capture macroeconomic tail risks using tail-focused scoring rules and comparisons with quantile regression (QR) benchmarks, following \citet{carriero2024capturing}. This separation allows us to distinguish improvements in overall density forecasting from gains specific to tail behavior.

We consider six specifications of the proposed level--volatility DFM, varying the number of level factors $J\in\{2,4\}$ and the number of common volatility factors $k\in\{1,2\}$. The baseline specification consists of two level factors and one common volatility factor $(J=2,k=1)$. For comparison, benchmark DFMs with the same number of level factors but without common volatility factors are also estimated. All specifications are estimated recursively using an expanding-window scheme and the same information set, ensuring a controlled comparison across models.

Forecast accuracy is evaluated using both point and density-based measures, following the standard predictive evaluation literature (e.g., \citealp{GneitingRaftery2007,geweke2010comparing}). Point forecasts are assessed using root mean squared forecast errors (RMSE), computed from predictive means. Density forecasts are evaluated using the continuous ranked probability score (CRPS), which provides a proper scoring rule for the full predictive distribution and is robust to non-Gaussian features such as asymmetry and fat tails. Both CRPS and log predictive scores are strictly proper scoring rules, ensuring that the evaluation rewards well-calibrated and sharp predictive distributions (see \citet{GneitingRaftery2007}).

In addition, to assess the ability of the model to capture macroeconomic tail risks, we consider tail-focused evaluation metrics that place greater weight on specific regions of the predictive distribution. In particular, we employ threshold-weighted versions of the CRPS, following \citet{gneiting2011comparing}, as well as comparisons with quantile-regression benchmarks following the Growth-at-Risk literature (e.g., \citealp{AdrianBoyarchenkoGiannone2019,carriero2024capturing}).\footnote{For implementation details and alternative representations of weighted scoring rules, see \citet{allen2024weighted}.} These measures allow us to evaluate whether the proposed framework improves the characterization of downside and upside risks beyond standard density metrics. A detailed discussion of these tail-focused measures is provided in the following subsection.

For output, employment, and inflation, which are modeled in first log-differences, forecast performance is evaluated in terms of cumulative growth rates over the forecast horizon.

\subsection{Forecasting results}

Table~\ref{tab:crps_rmse_comparison} reports forecast accuracy for the competing specifications at the one-year ($h=4$) and two-year ($h=8$) horizons. Panel~A presents results for the full evaluation sample (1992Q2--2025Q2), while Panel~B reports results for the pre-COVID subsample (1992Q2--2019Q4). The The DFM (2F) row is reported in absolute loss levels, while all remaining entries are expressed relative to DFM (2F). Values below one indicate improved forecast accuracy. Statistical significance is assessed using one-sided Diebold--Mariano tests based on CRPS loss differentials. While Panel~A is informative about model performance during periods of heightened macroeconomic stress, Panel~B provides the cleaner baseline comparison for assessing systematic differences across specifications.
Across both samples, gains from the level--volatility specification are concentrated in density forecasts. In Panel~A, specifications that incorporate common volatility factors deliver substantial and statistically significant improvements for employment, inflation, and the federal funds rate. The gains are particularly pronounced for employment, where benchmark DFMs exhibit very large relative losses during extreme labor-market episodes. The extremely large relative losses for employment reflect a small number of extreme forecast errors during periods of abrupt labor market adjustment, particularly during the pandemic. As shown in Panel~B, these differences are substantially attenuated when excluding this period. In this setting, level--volatility models reduce forecast losses markedly across all specifications, indicating improved robustness to large and coordinated shocks rather than uniformly superior performance across all states of the economy.

For GDP growth, the evidence in Panel~A is more heterogeneous. Specifications with two level factors do not systematically outperform the benchmark, whereas models that combine four level factors with volatility factors deliver improvements, particularly at longer horizons. This pattern suggests that the contribution of volatility dynamics becomes more relevant once the conditional mean is sufficiently well captured.

Panel~B shows that these findings are not driven by the inclusion of the pandemic period. When the evaluation is restricted to the pre-COVID sample, improvements in density forecasts remain robust for employment, inflation, and the federal funds rate. In particular, level--volatility specifications continue to deliver statistically significant gains for employment across horizons, as well as consistent improvements for inflation and the policy rate. For GDP growth, improvements are more modest and primarily concentrated in specifications with richer factor structures.

Comparing Panels~A and~B indicates that the pandemic period amplifies differences in forecast performance—most notably for employment—but does not alter the qualitative ranking across models. In both samples, specifications with common volatility factors systematically outperform benchmark DFMs without such factors.

Turning to point forecasts, differences across models are limited in both panels. RMSE ratios are generally close to one for GDP growth, employment, and inflation, indicating that incorporating common volatility factors has little effect on the conditional mean. Small improvements are observed for inflation in some specifications, while differences for GDP and employment are negligible. For the federal funds rate, RMSE is slightly higher across most specifications, although the deterioration is economically small.

Overall, the results indicate that allowing for joint level--volatility dynamics improves the accuracy of predictive densities, while leaving point forecasts largely unchanged. This pattern is consistent with a framework in which volatility primarily affects the dispersion and shape of the predictive distribution rather than its central tendency.

\clearpage

\begin{table}[!p]
\centering
\caption{Forecast accuracy: CRPS and RMSE}
\label{tab:crps_rmse_comparison}
{\footnotesize
\setlength{\tabcolsep}{3.5pt}
\renewcommand{\arraystretch}{1.1}
\begin{tabular}{@{}lcccccccc@{}}
\toprule
& \multicolumn{2}{c}{GDP} & \multicolumn{2}{c}{EMP} & \multicolumn{2}{c}{CPI} & \multicolumn{2}{c}{FFR} \\
& 1y & 2y & 1y & 2y & 1y & 2y & 1y & 2y \\
\midrule
\multicolumn{9}{@{}l}{\textbf{Panel A: Full sample (1992Q2--2025Q2)}}\\
\midrule
\multicolumn{9}{@{}l}{\textit{CRPS}} \\
DFM (2F)       & 0.011 & 0.017 & 0.491 & 0.925 & 0.005 & 0.006 & 0.396 & 0.407 \\
DFM (4F)       & 1.943 & 2.023 & 1.194 & 1.016 & 1.095 & 1.103 & 1.162 & 1.142 \\
LV-DFM (2L,1V) & 1.753 & 2.552 & 0.017*** & 0.018*** & 0.751** & 0.744** & 0.675*** & 0.656*** \\
LV-DFM (2L,2V) & 1.849 & 2.621 & 0.022 & 0.021 & 0.759** & 0.744** & 0.691*** & 0.672*** \\
LV-DFM (4L,1V) & 0.919 & 0.965 & \textbf{0.016***} & \textbf{0.017***} & 0.751* & \textbf{0.737**} & \textbf{0.619***} & 0.592*** \\
LV-DFM (4L,2V) & \textbf{0.803***} & \textbf{0.781**} & 0.021 & 0.021 & \textbf{0.749*} & 0.740* & 0.621*** & \textbf{0.590***} \\
\addlinespace
\multicolumn{9}{@{}l}{\textit{RMSE}} \\
DFM (2F)       & 0.018 & 0.026 & 3.252 & 3.778 & 0.007 & 0.008 & 0.438 & 0.428 \\
DFM (4F)       & 2.091 & 3.131 & 1.989 & 2.936 & 1.052 & 1.026 & 1.121 & 1.068 \\
LV-DFM (2L,1V) & 1.085 & 1.282 & 0.623 & 0.843 & \textbf{0.953} & 1.045 & 1.075 & 1.043 \\
LV-DFM (2L,2V) & 1.797 & 1.309 & 0.674 & 0.904 & \textbf{0.953} & 1.051 & 1.105 & 1.053 \\
LV-DFM (4L,1V) & 1.030 & 0.999 & \textbf{0.605} & \textbf{0.807} & 0.999 & 1.002 & 1.028 & 1.009 \\
LV-DFM (4L,2V) & 1.005 & \textbf{0.993} & 0.607 & 0.821 & 0.995 & \textbf{0.973} & 1.034 & 1.016 \\
\midrule
\multicolumn{9}{@{}l}{\textbf{Panel B: Pre-COVID sample (1992Q2--2019Q4)}}\\
\midrule
\multicolumn{9}{@{}l}{\textit{CRPS}} \\
DFM (2F)       & 0.010 & 0.015 & 0.011 & 0.021 & 0.005 & 0.006 & 0.388 & 0.405 \\
DFM (4F)       & 0.971 & 1.000 & 1.012 & 1.010 & 1.184 & 1.346 & 1.064 & 1.044 \\
LV-DFM (2L,1V) & 0.998 & \textbf{0.900} & \textbf{0.707***} & \textbf{0.803*} & 0.696** & 0.724** & 0.665*** & 0.648*** \\
LV-DFM (2L,2V) & 0.992 & 0.908 & 0.942** & 0.959 & 0.724** & 0.761** & 0.681*** & 0.672*** \\
LV-DFM (4L,1V) & 0.932 & 0.961 & \textbf{0.676***} & \textbf{0.779**} & 0.670** & 0.683** & \textbf{0.602***} & 0.580*** \\
LV-DFM (4L,2V) & \textbf{0.923*} & 0.932 & 0.892*** & 0.932** & \textbf{0.664**} & \textbf{0.679**} & \textbf{0.602***} & \textbf{0.579***} \\
\addlinespace
\multicolumn{9}{@{}l}{\textit{RMSE}} \\
DFM (2F)       & 0.018 & 0.027 & 0.021 & 0.033 & 0.007 & 0.007 & 0.398 & 0.410 \\
DFM (4F)       & \textbf{0.936**} & \textbf{0.894**} & 0.991 & 0.980 & 1.044 & 1.015 & 1.101 & 1.060 \\
LV-DFM (2L,1V) & 1.080 & 1.116 & 0.972 & 1.000 & 0.926** & 1.073 & 1.131 & 1.052 \\
LV-DFM (2L,2V) & 1.070 & 1.114 & 0.996 & 1.031 & 0.953 & 1.144 & 1.161 & 1.078 \\
LV-DFM (4L,1V) & 1.006 & 1.008 & \textbf{0.950*} & \textbf{0.968} & 0.881 & \textbf{0.908} & 1.059 & 1.013 \\
LV-DFM (4L,2V) & 1.009 & \textbf{0.994} & 0.968 & 0.989 & \textbf{0.872} & \textbf{0.908} & 1.065 & 1.019 \\
\bottomrule
\end{tabular}
}
\vspace{2mm}
\noindent\begin{minipage}{\textwidth}
\footnotesize
\textit{Notes:} The The DFM (2F) row is reported in absolute loss levels, while all remaining entries are expressed relative to DFM (2F). Values below one indicate improved forecast accuracy. Bold entries denote the best-performing specification among models that outperform the benchmark. For output, employment, and inflation, which are specified in first log-differences, forecast performance is evaluated in terms of cumulative growth rates. CRPS denotes the continuous ranked probability score and RMSE is computed from predictive means. Asterisks denote one-sided Diebold--Mariano tests of equal predictive accuracy relative to DFM (2F): *** $p<0.01$, ** $p<0.05$, * $p<0.10$. GDP: real output; EMP: employment; CPI: consumer price inflation; FFR: federal funds rate.
\end{minipage}
\end{table}

\clearpage

\subsection{Tail Risk in Predictive Distributions}

\begin{figure}[!htbp]
 \centering
\includegraphics[width=\textwidth]{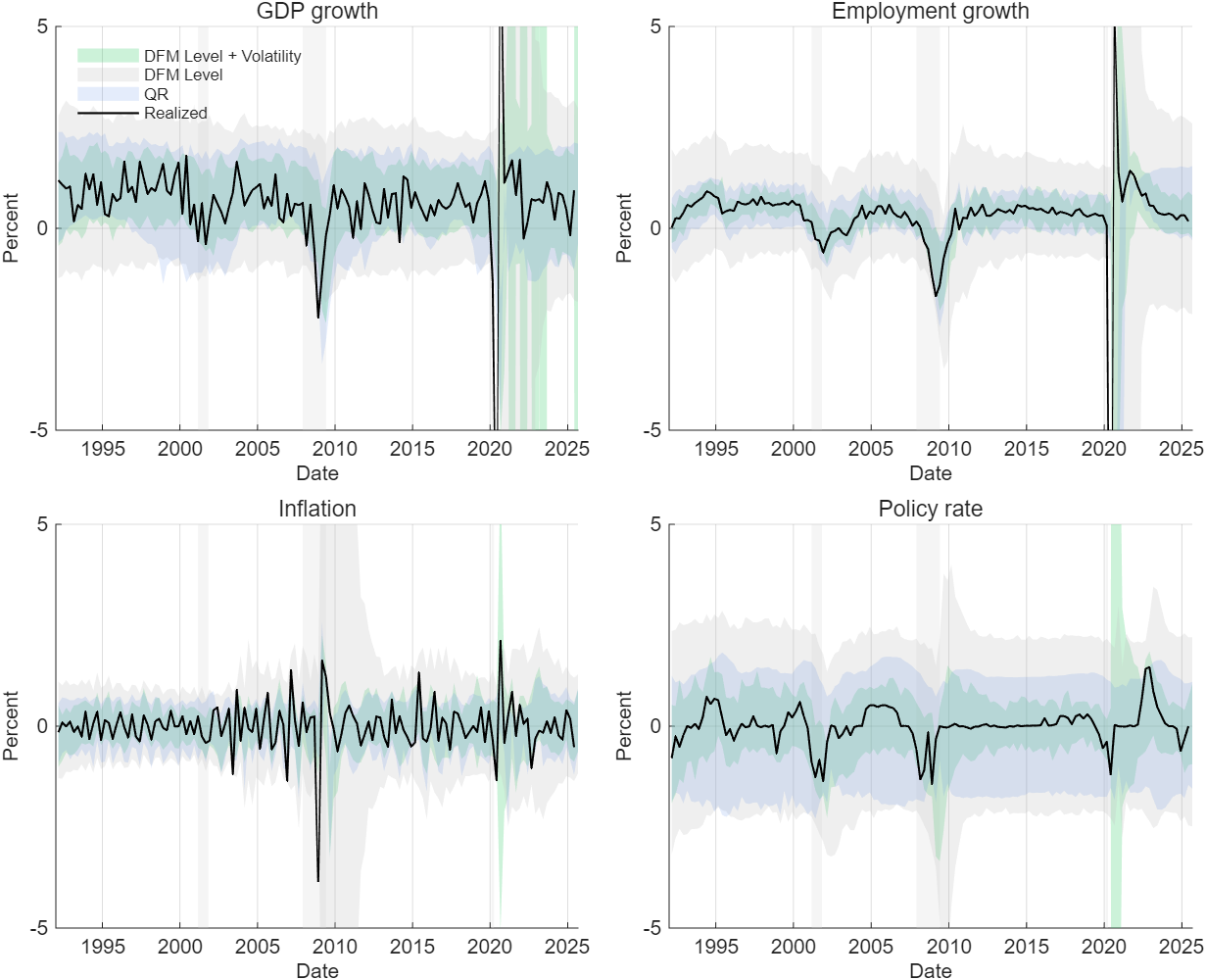}
\caption{One-period-ahead predictive quantiles}
\caption*{\footnotesize \textbf{Notes:} The figure reports one-quarter-ahead predictive 5\% and 95\% quantiles for selected macroeconomic variables. Green lines correspond to the level--volatility DFM, grey lines correspond to the benchmark DFM without a common volatility factor, and blue dashed lines correspond to the quantile regression (QR) benchmark. For GDP growth, the QR specification follows the Growth-at-Risk regression used in \citet{carriero2024capturing}. The black line denotes realized outcomes.}
 \label{fig:percentiles}
\end{figure}

Figure~\ref{fig:percentiles} reports one-quarter-ahead predictive quantiles for key macroeconomic variables. The figure illustrates how the proposed level--volatility DFM generates time variation in macroeconomic tail risks. For each series, we plot realized outcomes together with the 5th and 95th percentiles of the predictive distribution from the level--volatility specification (green) and from a benchmark DFM without a common volatility factor (grey).

To provide a direct comparison with the Growth-at-Risk literature, we also report quantile regression (QR) estimates (blue). For GDP growth, the QR specification conditions on financial and macroeconomic predictors following \citet{carriero2024capturing} within the framework of \citet{AdrianBoyarchenkoGiannone2019}, while for the remaining variables QR is implemented as a univariate autoregressive quantile regression. Unlike QR, which directly targets specific quantiles, the level--volatility DFM generates tail behavior as an implication of a fully specified predictive density. The key difference is that in our framework tail risks arise endogenously from the interaction between level and volatility factors, rather than being imposed through reduced-form quantile regressions.

The relevant tail is variable-specific. For GDP growth, the focus is on the lower tail, reflecting downside risks to real activity. For inflation, the emphasis is on the upper tail, which captures risks of elevated inflation outcomes. In the proposed framework, these tail dynamics arise endogenously from the interaction between level and volatility factors, allowing shifts in uncertainty to affect both the dispersion and the location of the predictive distribution.

The level--volatility specification generates predictive distributions that adjust more strongly during episodes of macroeconomic stress. In downturns, declines in the level factor are accompanied by increases in the volatility factor, shifting the predictive distribution leftward and widening it asymmetrically. For GDP growth, this mechanism leads to a disproportionate expansion of the lower tail, consistent with Growth-at-Risk dynamics. For inflation, periods of elevated uncertainty are associated with a widening of the upper tail, reflecting greater upside risk. Figure~\ref{fig:percentiles} therefore provides direct visual evidence of this mechanism: during stress episodes, the level--volatility specification simultaneously shifts and widens the predictive distribution in the relevant tail, whereas the benchmark DFM does not capture this joint adjustment.

Table~\ref{tab:tailrisk_twcrps} evaluates tail forecast accuracy using threshold-weighted CRPS measures \citep{gneiting2011comparing}, which place explicit weight on the relevant regions of the predictive distribution. Results are reported for both the full sample and the pre-COVID period. QR benchmarks are shown in levels, while all other entries are expressed relative to QR, so that values below one indicate improved tail accuracy.

Several results emerge. First, benchmark DFMs without common volatility factors perform poorly in the tails, particularly for employment and inflation, indicating that models without a common volatility component fail to capture coordinated shifts in uncertainty that are central for tail dynamics.

Second, level--volatility specifications deliver large and systematic improvements relative to benchmark DFMs across all variables and horizons. These gains are strongest in variables where tail risks are tightly linked to aggregate uncertainty.

Third, while QR remains a strong benchmark---particularly for GDP downside risk---the level--volatility DFM performs competitively and in some cases outperforms QR. Importantly, the proposed model delivers these improvements within a coherent joint predictive distribution, allowing for a unified characterization of central and tail risks. This contrasts with QR approaches, which model quantiles directly but abstract from the interaction between conditional means and variances.

Finally, results are broadly similar when the evaluation is restricted to the pre-COVID sample (Panel~B), indicating that the improvements are not driven solely by extreme pandemic observations. The relative ranking across models remains stable, with level--volatility specifications consistently outperforming standard DFMs in the tails.

Overall, the evidence indicates that modeling the joint dynamics of level and volatility factors substantially improves the characterization of macroeconomic tail risks. These gains are strongest for variables where tail behavior is tightly linked to aggregate uncertainty, especially in the lower tail of GDP growth and the upper tail of inflation, and remain meaningful even where QR provides a demanding benchmark.

\clearpage
\begin{table}[!htbp]
\centering
\caption{Forecast accuracy: predictive performance in the tails}
\label{tab:tailrisk_twcrps}
{\footnotesize
\setlength{\tabcolsep}{3pt}
\renewcommand{\arraystretch}{1.04}

\noindent\textbf{Panel A: Full sample (1992Q2--2025Q2)}\par\vspace{0.5mm}
\begin{tabular}{@{}lcccccccc@{}}
\toprule
& \multicolumn{2}{c}{GDP} & \multicolumn{2}{c}{EMP} & \multicolumn{2}{c}{CPI} & \multicolumn{2}{c}{FFR} \\
& 1y & 2y & 1y & 2y & 1y & 2y & 1y & 2y \\
\midrule

\multicolumn{9}{@{}l}{\textit{twCRPS (left tail)}} \\
\midrule
QR              & 0.004 & 0.003 & 0.006 & 0.008 & 0.002 & 0.003 & 0.122 & 0.117 \\
DFM (2F)        & 0.750*** & 1.000*** & 44.833*** & 62.250*** & 1.500*** & 1.000*** & 1.672*** & 1.803*** \\
DFM (4F)        & 2.250*** & 4.000*** & 42.167*** & 50.250*** & 1.500*** & 1.333*** & 1.926*** & 2.043*** \\
LV-DFM (2L,1V)  & 1.250*** & 2.667*** & \textbf{0.833***} & \textbf{0.875***} & 1.000*** & \textbf{0.667***} & 1.156*** & 1.197*** \\
LV-DFM (2L,2V)  & 1.250*** & 2.667*** & \textbf{0.833***} & \textbf{0.875***} & 1.000*** & \textbf{0.667***} & 1.180*** & 1.231*** \\
LV-DFM (4L,1V)  & \textbf{0.750***} & \textbf{0.667***} & \textbf{0.833***} & \textbf{0.875***} & 1.000*** & \textbf{0.667***} & 1.041*** & 1.085*** \\
LV-DFM (4L,2V)  & \textbf{0.750***} & \textbf{0.667***} & \textbf{0.833***} & \textbf{0.875***} & 1.000*** & \textbf{0.667***} & 1.041*** & 1.085*** \\

\addlinespace
\multicolumn{9}{@{}l}{\textit{twCRPS (right tail)}} \\
\midrule
QR              & 0.016 & 0.041 & 0.009 & 0.022 & 0.002 & 0.002 & 0.116 & 0.101 \\
DFM (2F)        & 0.500*** & 0.341*** & 24.667*** & 19.409*** & 1.000*** & 1.500*** & 1.655*** & 1.941*** \\
DFM (4F)        & 0.812*** & 0.537*** & 37.111*** & 24.409*** & 1.500*** & 2.000*** & 1.940*** & 2.228*** \\
LV-DFM (2L,1V)  & 0.625*** & 0.488*** & 0.333*** & 0.409*** & 1.000*** & 1.000*** & 1.086*** & 1.257*** \\
LV-DFM (2L,2V)  & 0.562*** & 0.463*** & 0.667*** & 0.591*** & 1.000*** & 1.000*** & 1.121*** & 1.287*** \\
LV-DFM (4L,1V)  & 0.438*** & 0.317*** & \textbf{0.333***} & \textbf{0.364***} & \textbf{0.500***} & 1.000*** & 1.017*** & 1.119*** \\
LV-DFM (4L,2V)  & \textbf{0.375***} & \textbf{0.293***} & 0.556*** & 0.545*** & \textbf{0.500***} & 1.000*** & \textbf{0.914***} & 1.059*** \\
\bottomrule
\end{tabular}

\vspace{2.5mm}

\noindent\textbf{Panel B: Pre-COVID sample (1992Q2--2019Q4)}\par\vspace{0.5mm}
\begin{tabular}{@{}lcccccccc@{}}
\toprule
& \multicolumn{2}{c}{GDP} & \multicolumn{2}{c}{EMP} & \multicolumn{2}{c}{CPI} & \multicolumn{2}{c}{FFR} \\
& 1y & 2y & 1y & 2y & 1y & 2y & 1y & 2y \\
\midrule

\multicolumn{9}{@{}l}{\textit{twCRPS (left tail)}} \\
\midrule
QR              & 0.004 & 0.003 & 0.006 & 0.009 & 0.002 & 0.002 & 0.123 & 0.118 \\
DFM (2F)        & 0.750*** & \textbf{0.667***} & 0.833*** & 0.889*** & 1.500*** & 1.500*** & 1.667*** & 1.814*** \\
DFM (4F)        & 0.750*** & \textbf{0.667***} & 0.833*** & 0.889*** & 1.500*** & 2.000*** & 1.748*** & 1.856*** \\
LV-DFM (2L,1V)  & 0.750*** & \textbf{0.667***} & 0.833*** & 0.889*** & 1.000*** & 1.000*** & 1.163*** & 1.203*** \\
LV-DFM (2L,2V)  & 0.750*** & 1.000*** & 0.833*** & \textbf{0.778***} & 1.000*** & 1.000*** & 1.187*** & 1.246*** \\
LV-DFM (4L,1V)  & 0.750*** & \textbf{0.667***} & 0.833*** & 0.889*** & 1.000*** & 1.000*** & 1.049*** & 1.093*** \\
LV-DFM (4L,2V)  & 0.750*** & \textbf{0.667***} & 0.833*** & \textbf{0.778***} & 1.000*** & 1.000*** & 1.041*** & 1.085*** \\

\addlinespace
\multicolumn{9}{@{}l}{\textit{twCRPS (right tail)}} \\
\midrule
QR              & 0.015 & 0.041 & 0.008 & 0.021 & 0.002 & 0.002 & 0.101 & 0.091 \\
DFM (2F)        & 0.467*** & 0.317*** & 0.750*** & 0.571*** & 1.000*** & 1.500*** & 1.812*** & 2.110*** \\
DFM (4F)        & 0.467*** & \textbf{0.293***} & 0.750*** & 0.619*** & 1.500*** & 2.000*** & 1.950*** & 2.242*** \\
LV-DFM (2L,1V)  & 0.467*** & 0.341*** & \textbf{0.375***} & 0.429*** & \textbf{0.500***} & 1.000*** & 1.139*** & 1.330*** \\
LV-DFM (2L,2V)  & 0.467*** & 0.341*** & 0.625*** & 0.571*** & 1.000*** & 1.000*** & 1.168*** & 1.385*** \\
LV-DFM (4L,1V)  & \textbf{0.400***} & \textbf{0.293***} & \textbf{0.375***} & \textbf{0.381***} & \textbf{0.500***} & 1.000*** & 1.040*** & 1.165*** \\
LV-DFM (4L,2V)  & \textbf{0.400***} & \textbf{0.293***} & 0.625*** & 0.571*** & \textbf{0.500***} & 1.000*** & 1.050*** & 1.176*** \\
\bottomrule
\end{tabular}

}
\vspace{1mm}
\noindent\begin{minipage}{\textwidth}
\footnotesize
\textit{Notes:} QR rows are reported in absolute loss levels, while all non-QR entries are expressed relative to QR. Values below one indicate improved forecast accuracy. Bold entries denote the best-performing specification among models that outperform the benchmark. For output, employment, and inflation, which are specified in first log-differences, forecast performance is evaluated in terms of cumulative growth rates. twCRPS denotes threshold-weighted continuous ranked probability scores following \citet{gneiting2011comparing}. Asterisks denote DM--West tests of equal predictive accuracy against the QR benchmark: *** $p<0.01$, ** $p<0.05$, * $p<0.10$. GDP: real output; EMP: employment; CPI: consumer price inflation; FFR: federal funds rate.
\end{minipage}
\end{table}

\clearpage
\section{\label{Decomp}Decomposing level and volatility drivers of inflation}
This section uses the proposed level--volatility DFM to conduct a semi-structural decomposition of international inflation dynamics into common and group-specific components operating through both conditional means and conditional volatilities. The exercise builds on the global inflation literature, which documents a substantial role for common international forces alongside pronounced cross-country heterogeneity in inflation outcomes \citep{MumtazSurico2012,HaKoseOhnsorge2023}. In particular, we follow the decomposition framework of \citet{Chernis2025}, who models inflation as the sum of a global component and regional components associated with advanced economies and emerging market and developing economies (EMDEs), and adapt it to a setting in which both inflation levels and inflation uncertainty are driven by latent common factors.

The empirical implementation uses quarterly CPI inflation data from the World Bank Global Inflation Database \citep{HaKoseOhnsorge2023}. Let $\pi_{i,t}$ denote quarterly CPI inflation in country $i$, for $i=1,\dots,N$, and define the group indicator
\[
s_i \equiv \mathbf{1}(i \in \text{AE}),
\]
which equals one for advanced economies and zero for EMDEs. Inflation is decomposed into a conditional mean component driven by global and regional level factors and an innovation whose conditional variance is governed by global and regional volatility factors.

The measurement equation for inflation is specified as
\begin{equation}
\pi_{i,t}
= b_i^{w} f_{w,t}
+ s_i\, b_i^{a} f_{a,t}
+ (1-s_i)\, b_i^{e} f_{e,t}
+ v_{i,t},
\label{eq:infl_meas_decomp}
\end{equation}
where $f_{w,t}$ denotes the global (world) level factor, and $f_{a,t}$ and $f_{e,t}$ denote regional level factors for advanced economies and EMDEs, respectively. The coefficients $b_i^{w}$, $b_i^{a}$, and $b_i^{e}$ govern country-specific exposure to the global and regional inflation components.

To allow for persistence in country-specific inflation dynamics, the idiosyncratic component follows an autoregressive process,
\begin{equation}
v_{i,t}
= \sum_{\ell=1}^{L_{\pi}} \rho_{i,\ell} v_{i,t-\ell}
+ \epsilon_{i,t},
\qquad
\epsilon_{i,t} \sim \mathcal{N}(0,r_{i,t}),
\label{eq:infl_idio_ar}
\end{equation}
where $r_{i,t}$ denotes the conditional variance of the idiosyncratic innovation.

Inflation uncertainty is modeled explicitly by allowing the conditional variance $r_{i,t}$ to depend on a set of latent global and regional volatility factors,
\begin{equation}
\log r_{i,t}
= \beta_i^{w}\, \mathcal{F}_{w,t}
+ s_i\, \beta_i^{a}\, \mathcal{F}_{a,t}
+ (1-s_i)\, \beta_i^{e}\, \mathcal{F}_{e,t}
- \log \lambda_{i,t}.
\label{eq:infl_vol_decomp}
\end{equation}
Here $\mathcal{F}_{w,t}$, $\mathcal{F}_{a,t}$, and $\mathcal{F}_{e,t}$ denote global and regional volatility factors, while $\lambda_{i,t}$ is a series-specific scale component that allows for heavy-tailed idiosyncratic shocks through the scale-mixture representation described in Section~\ref{empirical}. This specification allows inflation uncertainty to co-move systematically across countries through common international and regional volatility drivers, rather than being absorbed into purely idiosyncratic disturbances.

In the remainder of this section, subsection~\ref{Decomp:prop} describes the estimated global and regional level and volatility factors, and subsection~\ref{Decomp:Crosscountry} examines cross-country heterogeneity in their contributions using forecast error variance decompositions.

\subsection{Properties of inflation level and volatility factors}\label{Decomp:prop}

\begin{figure}[!htbp]
    \centering
    \begin{subfigure}[b]{\textwidth}
        \centering
        \includegraphics[width=0.9\textwidth]{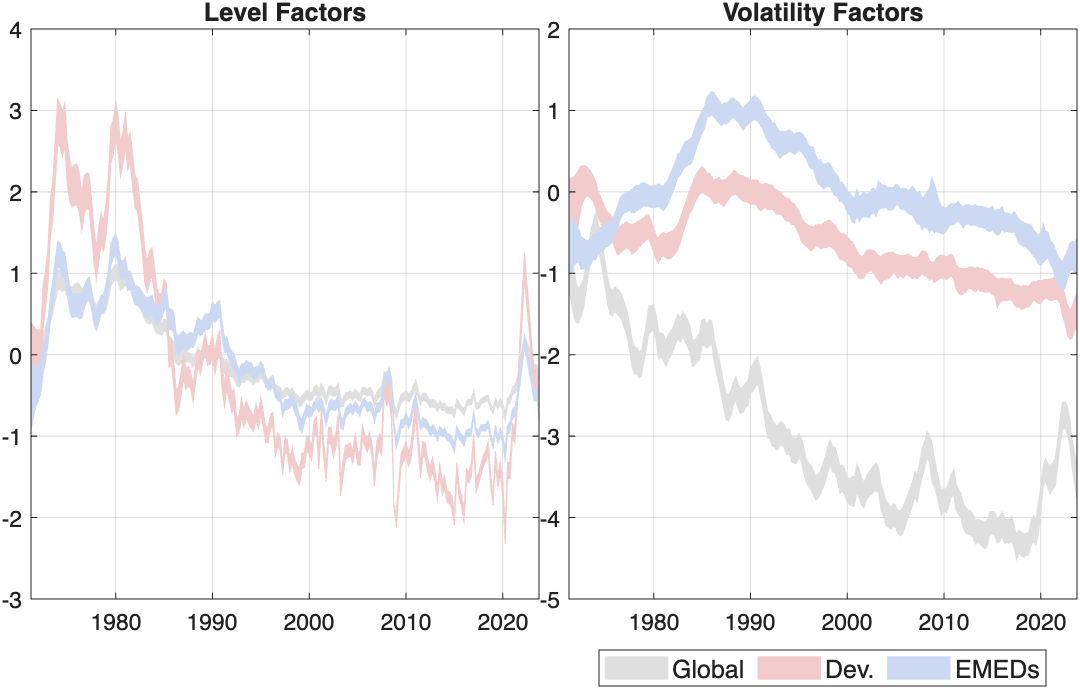}
        \caption{Dynamics of Level factors and volatility factors}
    \end{subfigure}

    \vspace{0.1cm}

    \begin{subfigure}[b]{\textwidth}
        \centering
        \includegraphics[width=0.99\textwidth]{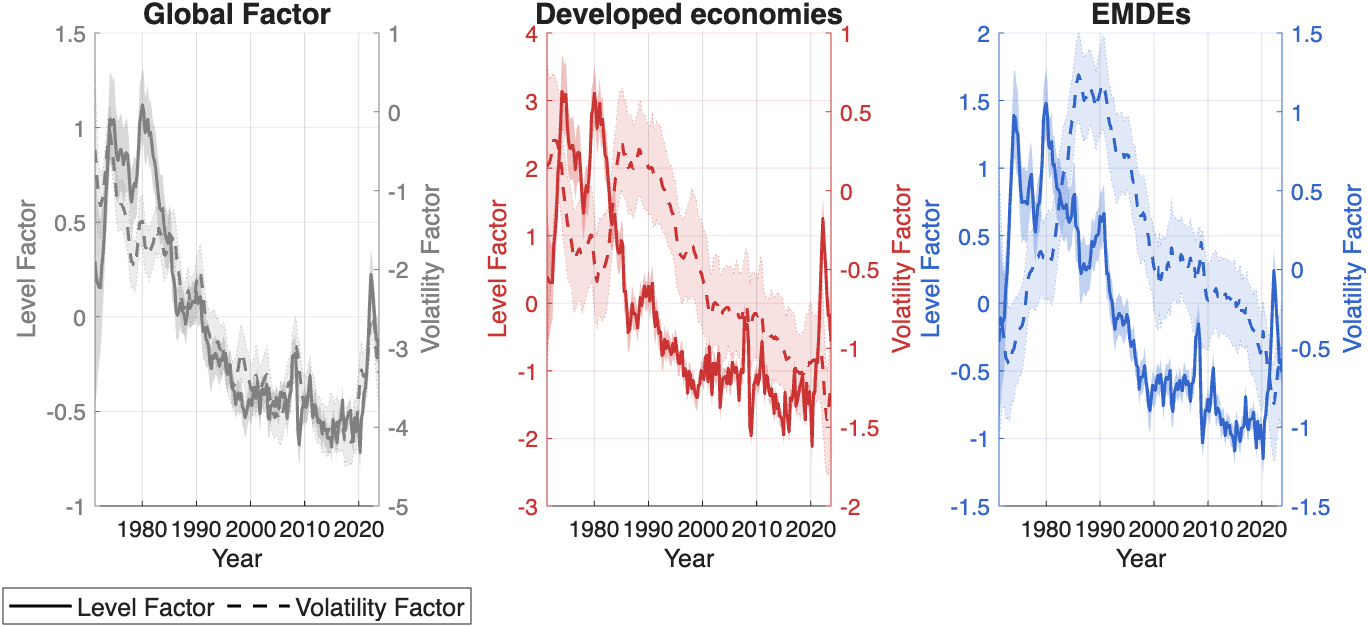}
        \caption{Global and regional drivers}
    \end{subfigure}

    \caption{Estimated level and volatility inflation factors: global and regional drivers}
    \label{fig:factors_region}
    \caption*{\footnotesize \textbf{Notes:} The figure displays posterior estimates of the level and volatility factors from the DFM. Panel (a) reports level factors (left) and volatility factors (right). Panel (b) overlays global and regional factors. Solid lines denote posterior medians for level factors and dashed lines denote posterior medians for volatility factors; shaded areas indicate 95\% credible intervals.}
\end{figure}

Figure~\ref{fig:factors_region} reports the estimated global and regional level and volatility factors. The level factors display pronounced low-frequency movements that are broadly synchronized across country groups. The global level factor captures major inflation episodes common across economies, including the high-inflation period of the 1970s, the subsequent disinflation, and the renewed increase in inflation at the end of the sample. Regional level factors for advanced economies and EMDEs exhibit similar long-run patterns but differ in timing and amplitude, consistent with heterogeneous transmission of global inflationary forces across country groups.

The volatility factors display distinct dynamics relative to the level factors. The global volatility factor declines over the 'Great Moderation' period and rises sharply toward the end of the sample, while regional volatility factors capture additional persistent group-specific movements. These patterns indicate that inflation uncertainty is shaped by both global and regional forces and that common variation in volatility is not a mechanical transformation of common variation in inflation levels.

Taken together, the estimates imply that international inflation dynamics are driven by common and group-specific components operating through both the conditional mean and the conditional volatility of inflation. Relative to standard three-factor decompositions, allowing for volatility factors isolates an additional dimension of commonality in inflation uncertainty that would otherwise be absorbed into idiosyncratic shocks.

\clearpage

\subsection{Country-level heterogeneity in inflation dynamics}\label{Decomp:Crosscountry}

\begin{figure}[!htbp]
    \centering
    \begin{subfigure}[b]{\textwidth}
        \centering
        \includegraphics[width=\textwidth]{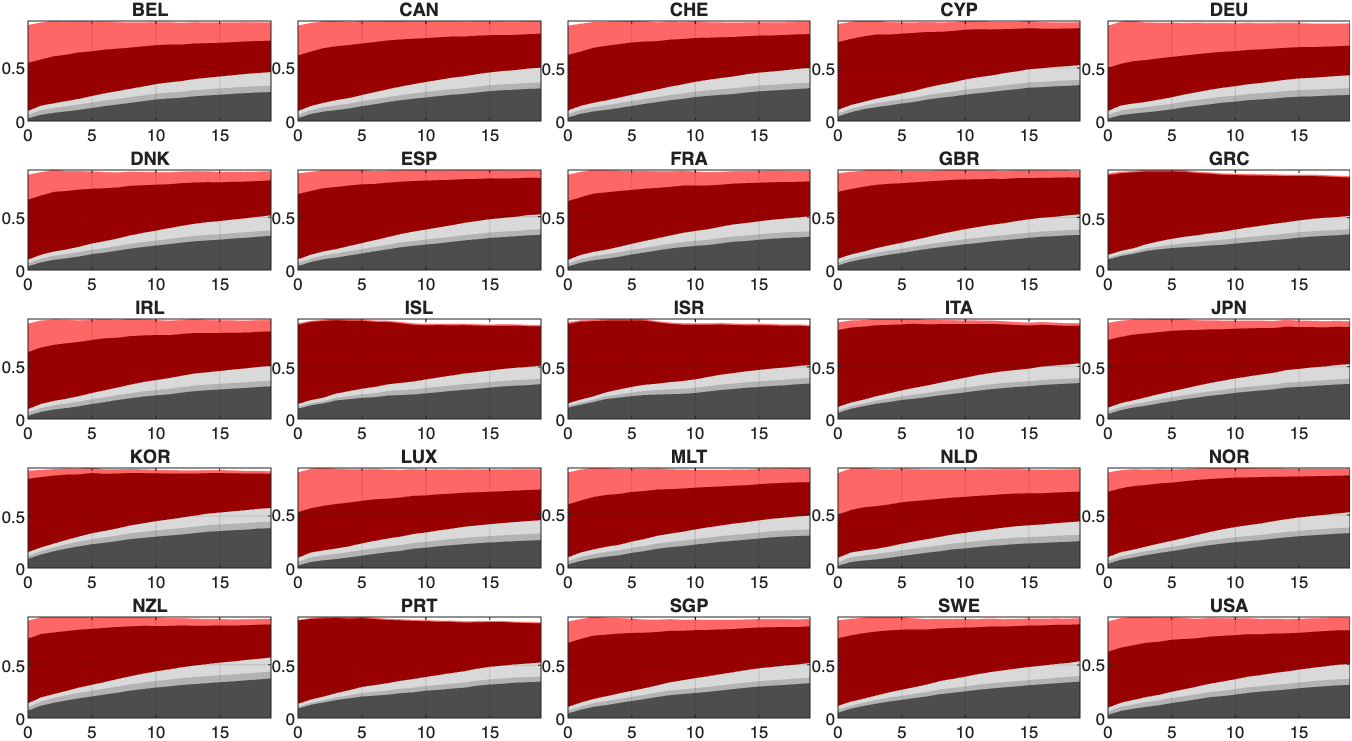}
        \caption{Developed economies}
        \label{fig:decomp_dev}
    \end{subfigure}

    \vspace{0.1cm}
        \begin{subfigure}[b]{\textwidth}
        \centering
        \includegraphics[width=\textwidth]{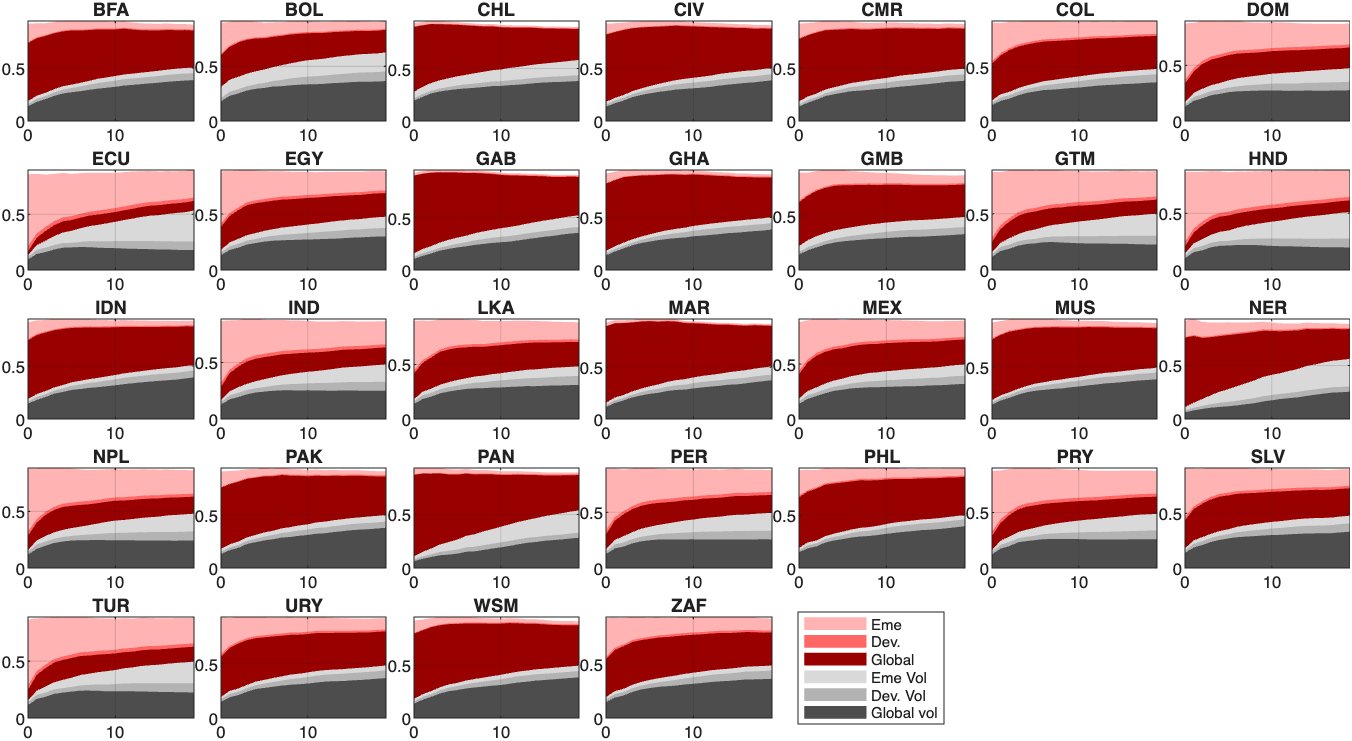}
        \caption{Emerging market and developing economies (EMDEs)}
        \label{fig:decomp_emde}
    \end{subfigure}

    \caption{Global and regional drivers of inflation: country-level FEVDs}
    \label{fig:decomp_combined}
    \caption*{\footnotesize \textbf{Notes:} The figure reports country-level cumulative forecast error variance decomposition (FEVD) shares for inflation. For each country, the shares are computed from impulse responses to one-standard-deviation shocks to the global, advanced-economy, and EMDE level and volatility innovations. The plotted objects are posterior medians of cumulative FEVD shares up to horizon $h$, averaged over time origins. Panel (a) shows developed economies and panel (b) shows EMDEs.}
\end{figure}

Figures~\ref{fig:decomp_dev} and \ref{fig:decomp_emde} report country-level cumulative forecast error variance decomposition (FEVD) profiles for inflation. For each posterior draw, we compute impulse responses to one-standard-deviation shocks to each of the six factor innovations—global, advanced-economy, and EMDE components in both levels and volatilities—and construct cumulative FEVD shares up to horizon $h$. The figures display posterior medians of these shares, averaged over time origins, so that each panel summarizes how the relative importance of global and regional level and volatility drivers evolves with the forecast horizon.

Two broad patterns emerge. In advanced economies, inflation dynamics are overwhelmingly driven by the global level factor at medium and long horizons. For virtually all countries in Figure~\ref{fig:decomp_dev}, the global level component quickly becomes the dominant contributor beyond short horizons and stabilizes at a high share of total variance. Regional (advanced-economy) level factors remain secondary: they are most visible at short horizons and fade relative to the global driver as the horizon increases.

At the same time, Figure~\ref{fig:decomp_dev} suggests that the strength of the regional advanced-economy component is not uniform within AEs. A subset of Northern European economies— notably Germany (DEU), the Netherlands (NLD), and Belgium (BEL)—exhibit a visibly larger advanced-economy regional share than, for example, the USA, Canada, Japan, or Korea, especially at intermediate horizons. A conservative interpretation is that within-AE inflation dynamics may retain an additional regional layer in parts of Europe, over and above the global component. \footnote{While this pattern could be consistent with differences in trade integration, energy exposure, or sectoral structure, the FEVDs alone do not identify the underlying mechanism.}

Volatility factors contribute a smaller but non-negligible share of inflation fluctuations in advanced economies. Across most AEs, the global volatility share rises gradually with the horizon, while regional volatility components remain comparatively modest. This horizon dependence is consistent with uncertainty acting as a persistent state variable that matters for medium-run predictability even if it is less important at very short horizons; importantly, the qualitative ranking of drivers is stable across AEs, with global level shocks first, followed by global volatility, and then regional components.

The picture is markedly different for EMDEs. Figure~\ref{fig:decomp_emde} shows substantially greater dispersion in both level and volatility contributions across countries. While the global level factor remains quantitatively important for many EMDEs, its dominance is far less uniform than in advanced economies. In a subset of countries—such as Indonesia (IDN), India (IND), and other emerging Asian economies—the global level component accounts for a large and persistent share of forecast error variance at medium and long horizons, resembling the pattern observed in AEs.

By contrast, several Latin American and Sub-Saharan African economies display a much stronger role for the EMDE regional level factor. In Latin America, Peru (PER), Panama (PAN), and Paraguay (PRY) exhibit a sizable and persistent contribution of the EMDE regional level component, in some cases remaining comparable to—or even exceeding—the global contribution at longer horizons. A similar pattern is visible for several African economies, including Ghana (GHA), The Gambia (GMB), and Niger (NER). This suggests that inflation dynamics in these economies are more tightly linked to region-specific forces than to the global inflation cycle alone.

Volatility drivers are also more prominent—and more heterogeneous—in EMDEs than in advanced economies. Importantly, this heterogeneity operates along two distinct dimensions. In some EMDEs, the global volatility factor gradually increases in importance with the horizon, indicating that worldwide uncertainty shocks shape medium-run inflation dynamics. In others—particularly several African and smaller open economies—the EMDE-specific volatility component accounts for a sizeable and persistent share of forecast error variance, pointing to common regional uncertainty shocks that are not captured by the global factor. Thus, it is not simply “uncertainty” in general that matters, but whether that uncertainty originates from global or EMDE-regional volatility forces.

Taken together, the FEVD evidence points to a layered structure of international inflation dynamics. Advanced economies are characterized by a highly dominant global mean component and relatively modest regional and volatility contributions. EMDEs, by contrast, display both stronger regional mean effects and a more substantial role for regional volatility shocks, alongside global drivers. This reinforces \cite{MumtazSurico2012}’s central insight that global inflation comovement coexists with cross-country heterogeneity, but extends it by showing that such heterogeneity is particularly pronounced along the volatility dimension and differs systematically between advanced and emerging economies.

\clearpage

\section{\label{conclu}Conclusion}

This paper develops a dynamic factor model in which common level and volatility factors evolve jointly within a unified state-space framework. Allowing conditional means and variances to interact endogenously embeds volatility-in-mean dynamics in a large-information setting and provides a tractable framework for characterizing macroeconomic tail risk.

The central insight is that when level and volatility factors follow a joint VAR, shocks to uncertainty propagate into both the dispersion and the location of the predictive distribution. This mechanism generates state-dependent asymmetries and time-varying downside risks without relying on extreme volatility realizations or ad hoc tail modeling.

The empirical results highlight the relevance of this structure. Incorporating common volatility factors yields systematic and statistically significant improvements in density forecast accuracy, particularly in the tails of the predictive distribution and at medium horizons, while effects on point forecasts are negligible. The model's value therefore lies in its ability to capture the evolution of macroeconomic risk rather than the conditional mean.

In the international inflation application, the joint framework reveals a layered structure of global inflation dynamics. Advanced economies are primarily driven by a global level factor, whereas emerging and developing economies exhibit stronger regional and volatility contributions. Cross-country heterogeneity is especially pronounced in the volatility dimension, which standard mean-based decompositions fail to capture.

Overall, the results show that jointly modeling level and volatility dynamics provides a tractable and economically meaningful framework for capturing macroeconomic uncertainty and its effects on the distribution of future outcomes. By allowing volatility to interact with the conditional mean through correlated factor dynamics, the model delivers a coherent characterization of macroeconomic tail risks and a unified alternative to reduced-form approaches such as quantile regressions. This underscores the importance of joint level--volatility dynamics for understanding and forecasting macroeconomic risk.

\appendix
\section{\label{app:model}Empirical model}
We consider the following state-space model:

\begin{eqnarray}
F_{t}
&=& c+\sum_{j=1}^{P}\beta _{j}F_{t-j}+e_{t}  \label{eq1_app} \\
e_{t} &\sim& N(0,\Omega)  \label{eq11_app} \\
F_{t}&=&\begin{pmatrix}
   f_{t} \\
    \mathcal{F}_{t}
\end{pmatrix}  \label{eq2_app} \\
r_{it} &=&\frac{e^{\mathcal{B}_{i}\mathcal{F}_{t}}}{\lambda_{it}}  \label{eq4_app} \\
\epsilon_{it}&\sim&N(0,r_{it}) \label{eq5_app} \\
v_{it}&=&\sum_{l=1}^{L}\rho_{i,l}v_{it-l}+\epsilon_{it} \label{eq51_app} \\
X_{it} &=&B_{i}f_{t}+v_{it}  \label{eq6_app}
\end{eqnarray}
where $i=1,2,\dots,N$ denotes the number of variables in the data set and $t=1,2,\dots,T$ denotes time.
\newline As in the FAVAR model of \cite{bernanke-boivin-eliasz-05}, each variable $X_{it}$ is related to a set of common factors $\underset{J\times1}{f_{t}}$ with loadings $B_{i}$. In a departure from the model of \cite{bernanke-boivin-eliasz-05}, the idiosyncratic components $v_{it}$ are autocorrelated, with persistence coefficients $\rho_{i}$, and heteroscedastic. The time-varying variance of the idiosyncratic shocks $\epsilon_{it}$ is denoted by $r_{it}$. This can be interpreted as a measure of uncertainty as it pertains to time-varying variance of the unpredictable component of $X_{it}$.
\newline It is well established that volatility displays co-movement across variables. Therefore, we allow the log of $r_{it}$ to depend on a set of common factors $\underset{k\times1}{\mathcal{F}_{t}}$ with loadings $\mathcal{B}_{i}$. Any idiosyncratic volatility movements are captured by $h_{it}=\frac{1}{\lambda_{it}}$. Higher frequency movements in the shocks (such as outliers) are captured by $
\lambda_{it}$. \citet{Geweke1993} shows
that assuming a Gamma prior for $\lambda_{it}$ of the form $
p\left( \lambda_{it}\right) =$ $\prod\limits_{m=1}^{T}\Gamma
\left( 1,\nu_{i }\right) $ leads to a scale mixture of normals for the $\epsilon_{it}$ and it follows a student
t-distribution with $\nu_{\lambda,i }$ degrees of freedom and time-varying
volatility given by $e^{\mathcal{B}_{i}\mathcal{F}_{t}}$ (see also \citealp{CHIU20171124}).

 This specification for $r_{it}$ is similar to that adopted by \cite{Carriero2018} and \cite{Efrem_sectoral}. However, unlike these papers we allow for contemporaneous and lagged interaction between levels and volatilities by postulating a VAR process for factors $F_{t}=\begin{pmatrix}
   f_{t} \\
    \mathcal{F}_{t}
\end{pmatrix} $. Thus, the model allows for volatility-in mean-dynamics as in \cite{MUMTAZ201810} and \cite{CaldaraSVOL}. Shocks to volatility factors can affect both the conditional volatility $r_{it}$ and the levels $X_{it}$

\subsection{Estimation}
We estimate the model using a Gibbs sampling algorithm. In this section we describe the prior distributions and sketch the steps of the algorithm.
\subsection{Priors and starting values}
\paragraph{VAR coefficients}
Let $\Gamma =vec\left( [\beta _{j}; c]\right) $. Following \cite{banbura2010large}, we employ a Minnesota type prior $N\left( \Gamma _{0},P_{0}\right) $. The prior mean for the
parameters on the first lag obtained by estimating individual AR(1)
regressions. We set the prior tightness for autoregressive coefficients $\tau =0.1$. A flat prior is used for the
intercept terms and the corresponding tightness is set equal to $c=1000$.
\paragraph{VAR error covariance $\Omega$}
We factorise the covariance as $\Omega=A^{-1} H A^{-1'}$ where $A$ is lower triangular and $H$ is diagonal. The non-zero and non-one elements of $A$ have Gaussian prior: $N(a_{0},v_{a})$. The diagonal elements of $H$ have an Inverse Gamma prior: $IG(h_{0},T_{h})$.
\paragraph{Factor loadings}
The prior for $\mathcal{B}_{i}$ and $B_i$ is Gaussian. To obtain the prior mean for former, we start with an initial estimate of the log idiosyncratic volatility $r_{it}$. This is obtained by running univariate stochastic volatility models on the idiosyncratic components $\hat{v}_{it}=X_{it}-B_{i,pca}F_{t,pca}$ where $pca$ denotes principal component (PCA) estimates of the factors and loadings. We then use PCA to extract factors and factor loadings from this initial estimate of $r_{it}$, where the latter are used as mean of $\mathcal{B}_{i}$. The prior mean for $B_i$ is set as $B_{i,pca}$. The covariance of the prior distributions is set to an identity matrix.
\paragraph{Other parameters}
We employ a hierarchical prior for elements of $\lambda_{it}$ (see
\cite{koop03})). The prior for $\lambda _{it}$ is a Gamma distribution $
\Gamma \left( 1,\nu_{i }\right) $. The degrees of freedom parameter $
\nu_{i }$ is treated as an unknown parameter with the prior: $\Gamma
\left( \nu _{0},2\right) $. We set $\nu_{0}=20$ in our
application.  The prior for $\rho_i$ is normal $N(\rho_0,V_{\rho})$.
\subsection{Gibbs algorithm}\label{app:gibbs}
The Gibbs algorithm samples from the following conditional posterior distributions in each iteration:
\begin{enumerate}
    \item $F(H|\Psi)$ where $\Psi$ denotes all other parameters and states. We write the orthogonal residuals of the VAR $F_{t}
= c+\sum_{j=1}^{P}\beta _{j}F_{t-j}+e_{t}$ as $Ae_t=u_t$. The $kth$ diagonal element of $H$, $h^k$ has an $IG$ posterior with scale parameter $u_{i}'u_{i}+h_0$ and degrees of freedom $T+T_{0}$
    \item $F(A|\Psi)$. Note that $Ae_t=u_t$ represents a system of linear equations where the variance of $u_t$ is $H$. The conditional distributions for a linear
regression apply to each equation of this system. The $kth$ equation of this
system is given as $e_{t}^{k}=-e_{t}^{-k}\alpha_k +u_{t}^{k}$ where the
superscript $k$ denotes the $kth$ column of the residual matrix while $-k$
denotes columns $1$ to $k-1$ and $\alpha_k$ denotes the non-zero and non-one elements of $A$ pertaining to the kth equation.
\item $F(\Gamma|\Psi)$. We sample the coefficients of the VAR in equation \ref{eq1} using the algorithm described in \cite{CARRIERO2022506} which allows the draws to be carried out equation by equation.
\item $F(B_i|\Psi)$. Given a draw of the factors $f_t$, equation \ref{eq6} represents a series of linear regressions with serially correlated and heteroscedastic residuals. Conditional on the serial correlation coefficients $\rho_{i,l}$ and the volatility $r_{it}$, each equation can be written as:
\begin{equation}
    \frac{X_{it}-\sum_{l=1}^{L}\rho_{i,l}X_{it-l}}{\sqrt{r_{it}}}=B_{i}\left(\frac{f_{t}-\sum_{l=1}^{L}\rho_{i,l}f_{t-l}}{\sqrt{r_{it}}}\right)+\tilde{v}_{it}
\end{equation}
where  $\tilde{v}_{it}\sim N(0,1)$

The conditional posterior for the coefficients of this transformed regression is Gaussian with well known formulas for the mean and variance (see for e.g. \cite{kim-nelson-98}).
\item $F(\rho_i|\Psi)$. Given the idiosyncratic components $v_{it}$, equation \ref{eq51} is a linear regression with heteroscedatic residuals $\epsilon_{it}$. As in the previous step a GLS step can be used to remove the heteroscedasticity (i.e. dividing both sides of the regression by $\sqrt{r_{it}}$). The standard formula for the conditional posterior for linear regression coefficient in a Gaussian model then applies to the transformed model.
\item $F(\mathcal{B}_i|\Psi)$ For each $i$ we draw the volatility factor loading $\mathcal{B}_i$ using a random walk Metropolis Hastings step. We draw a candidate loading from $\mathcal{B}_{i}^{new}\sim N(\mathcal{B}_{i}^{old},V_{\mathcal{B}})$. Given $\mathcal{F}_t$ and $h_{it}$ the implied value of the volatility $r_{it}^{new}$ can be easily constructed using equation \ref{eq4}. Conditional on the remaining parameters and the factors $f_t$, equation \ref{eq6} constitutes a linear regression with a known form of heteroscedasticity and serial correlation and one can easily evaluate the Gaussian log-likelihood after a GLS transformation and construct the value of the posterior at the candidate draw. The posterior can be evaluated at the previous draw using the same procedure with the previous value of the loading $\mathcal{B}_i^{old}$ and the acceptance probability can be calculated. We calibrate $V_{\mathcal{B}}$ to keep the acceptance rates between $20$ \% and $40$ \%.

\item $F(\lambda_{it}|\Psi)$: As described in \cite{koop03}, the conditional posterior of $\lambda
_{it}$ is a Gamma distribution with the following mean and
degrees of freedom:
\begin{equation*}
m=\left( \nu _{i}+1\right) /\frac{1}{\tilde{r}_{it}}\epsilon_{it}^{2}+\nu _{i}
\end{equation*}

\begin{equation*}
df=\nu _{i}+1
\end{equation*}
where $\tilde{r}_{it}=e^{\mathcal{B}_{i}\mathcal{F}_{t}}$

\item $F(\nu_{i}|\Psi)$ This conditional distribution is non-standard and is given by:
\begin{equation*}
G\left( \nu _{i}\right) \propto \left( \frac{\nu _{i}}{2}\right) ^{\frac{
T\nu _{i}}{2}}\Gamma \left( \frac{\nu _{i}}{2}\right) ^{-T}\exp \left(
-\left( \frac{1}{\nu _{0}}+0.5\sum\limits_{t=1}^{T}\left[ \ln \left(
-\lambda _{i,t}^{-1}\right) +\lambda _{i,t}\right] \right) \nu _{i}\right)
\end{equation*}
We use a random walk Metropolis Hastings algorithm to draw from this
conditional posterior distribution.

\item $F(F_t|\Psi)$. Conditional on the remaining parameters, the model can be written in state-space form. The observation equations are:
\begin{eqnarray}
    X_{it}-\sum_{l=1}^{L}\rho_{i,l}X_{it-l}&=&B_i\left(f_t-\sum_{l=1}^{L}\rho_{i,l}f_{t-l}   \right)+\epsilon_{it} \\
    \epsilon_{it}&=&\left(\frac{e^{\mathcal{B}_{i}\mathcal{F}_{t}}}{\lambda_{it}} \right)^{\frac{1}{2}}\eta_{it},\eta_{it}\sim N(0,1)
\end{eqnarray}
The transition equation is given by equation \ref{eq1}. Given the non-linearity in the second observation equation, we draw the states using the particle Gibbs sampler with ancestor sampling introduced by \cite{JMLR:v15:lindsten14a}. We use the version of their sampler that allows for multiple lags in the transition equation and hence accounts for a degenerate transition density.
\end{enumerate}

\section{Simulation Design and Calibration}\label{app:MCsim}

To assess finite-sample performance, we conduct a Monte Carlo experiment in which data are generated from the state-space system described in Appendix~\ref{app:model}. The calibration is designed to resemble a medium-sized macroeconomic panel with persistent level and volatility dynamics and cross-sectional heterogeneity.

\paragraph{Model dimensions and sample size.}
The cross-sectional dimension is set to $N=100$. The latent state vector is $F_t=(f_t,\mathcal{F}_t)'$, comprising one level factor ($J=1$) and one volatility factor ($K_v=1$). We generate $T=600$ observations and discard the first 100 as burn-in, leaving an estimation sample of $T=500$.

\paragraph{Factor dynamics.}
The joint dynamics of level and volatility factors follow a VAR(1):
\begin{equation}
F_t = \Phi F_{t-1} + e_t, \qquad e_t \sim \mathcal{N}(0,\Sigma),
\end{equation}
with
\begin{equation}
\Phi =
\begin{pmatrix}
0.9 & -0.1 \\
0.1 & 0.9
\end{pmatrix},
\qquad
\Sigma =
\begin{pmatrix}
0.2 & 0.02 \\
0.02 & 0.2
\end{pmatrix}.
\end{equation}
This calibration implies highly persistent factors and allows for dynamic interaction between level and volatility innovations.

\paragraph{Idiosyncratic components and heavy tails.}
The idiosyncratic component follows an AR(1):
\begin{equation}
v_{it} = \rho_i v_{i,t-1} + \epsilon_{it},
\end{equation}
where $\rho_i \sim U[0.1,0.7]$ independently across $i$. Innovations are generated from a Student-$t$ distribution with degrees of freedom $\nu_i \sim U\{10,\dots,30\}$ to introduce cross-sectional heterogeneity in tail thickness. Shocks are scaled by the time-varying variance
\[
r_{it}=\exp(\mathcal{B}_i \mathcal{F}_t),
\]
so that volatility co-moves across series through the common volatility factor.

\paragraph{Loadings and identification.}
The loading matrices for the level factors ($B_i$) and volatility factors ($\mathcal{B}_i$) are drawn from a standard normal distribution. To ensure structural identification, the first $K \times K$ and $K_v \times K_v$ blocks of the loading matrices are restricted to an identity matrix.

\begin{figure}[!htbp]
 \centering
 \caption{Latent Factors: Estimated and True}
 \label{fig:factors}
 \includegraphics[width=\textwidth]{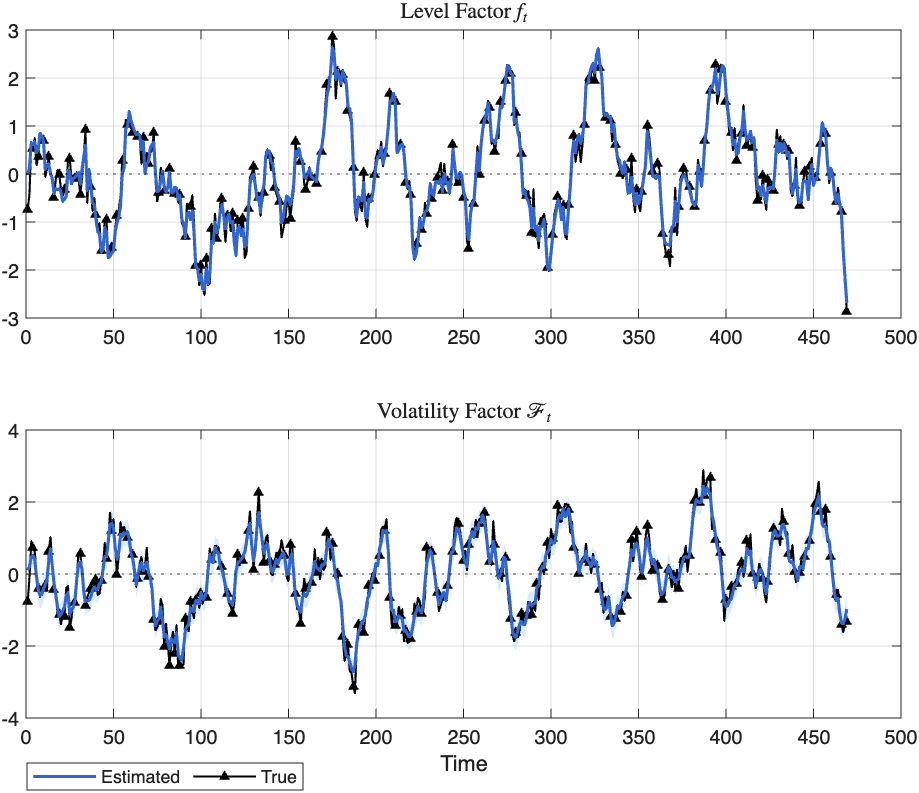}
 \caption*{\footnotesize Notes: The figure compares posterior median estimates of the latent factors with their true simulated values. Shaded areas denote 90\% credible bands. Factors are shown under the standard normalization used in estimation, which fixes scale and sign for comparability. The top panel reports the level factor ($f_t$), while the bottom panel reports the volatility factor ($\mathcal{F}_t$).}
\end{figure}

\begin{figure}[!htbp]
 \centering
 \begin{subfigure}[b]{\textwidth}
  \centering
  \includegraphics[width=\textwidth]{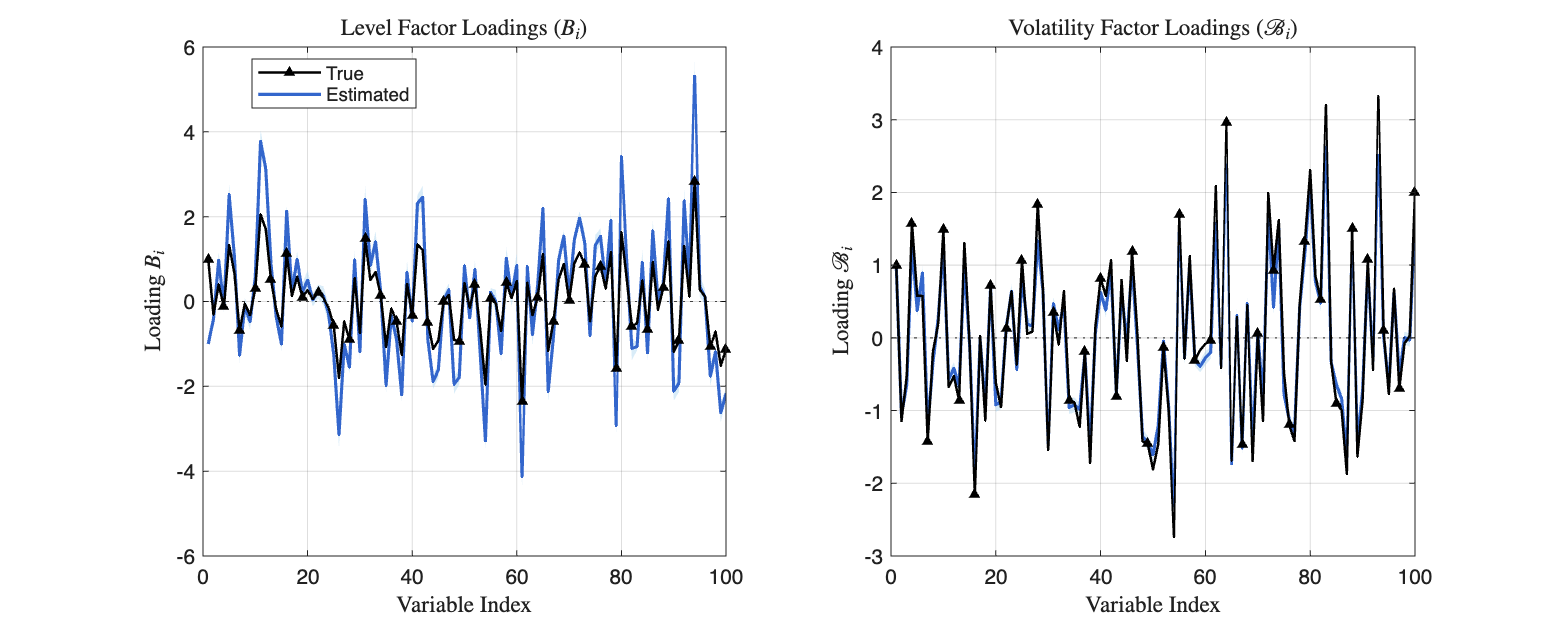}
  \caption{Factor Loadings}
  \label{fig:scatter}
 \end{subfigure}
 \vspace{0.2cm}
 \begin{subfigure}[b]{\textwidth}
  \centering
  \includegraphics[width=0.65\textwidth]{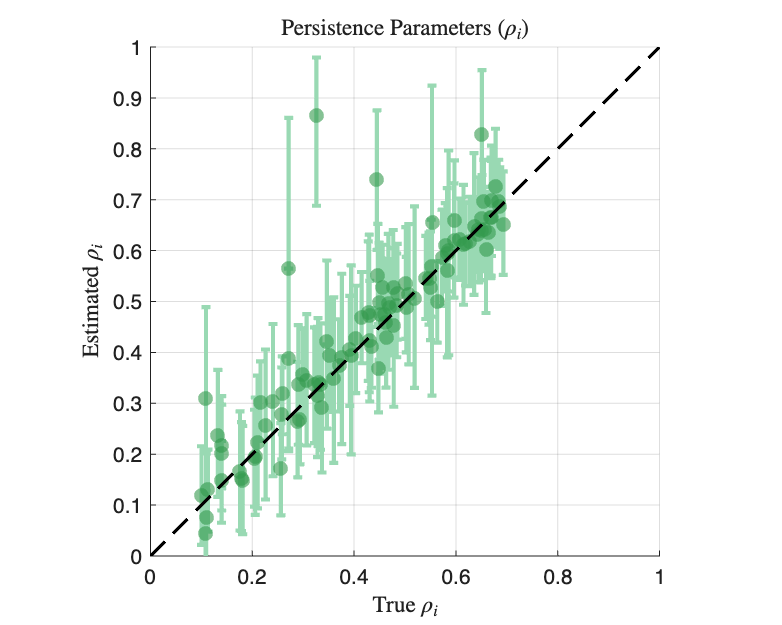}
  \caption{Persistence Parameters}
  \label{fig:persistence_scatter}
 \end{subfigure}
 \caption{Estimated versus True: Loadings and Persistence Parameters}
 \caption*{\footnotesize Notes: The figure displays scatter plots comparing true parameter values (horizontal axis) against posterior median estimates (vertical axis). Panel (a) reports factor loadings for the level factor ($B_i$) and volatility factor ($\mathcal{B}_i$). Panel (b) reports AR(1) persistence parameters ($\rho_i$). Vertical bars denote 90\% credible intervals. The dashed 45-degree line represents perfect recovery.}
\end{figure}

Figure~\ref{fig:factors} reports posterior inference for the latent level and volatility factors. Posterior medians closely track the true simulated paths over the sample, and the associated 90\% credible bands remain tight around the median. Both factors are accurately recovered throughout the sample.At the cross-sectional level, Figure~\ref{fig:scatter} plots the true and estimated factor loadings across variables. The estimated loadings closely follow the true cross-sectional patterns for both the level and volatility factors, indicating that the estimator successfully recovers the underlying heterogeneity across variables. For the volatility factor, the estimated loadings are very close to the true values, with the two series largely overlapping across the cross-section. For the level factor, the estimated loadings reproduce the same cross-sectional pattern, with small differences in overall scale relative to the true loadings.

Figure~\ref{fig:persistence_scatter} reports posterior inference for the idiosyncratic persistence parameters in scatter plot format. Posterior medians lie close to the true values and cluster around the 45-degree line for the majority of series, indicating accurate recovery.

\begin{figure}[!htbp]
 \centering
 \caption{Estimated versus True: Degrees of Freedom}
 \label{fig:dof_timeseries}
 \includegraphics[width=\textwidth,height=0.30\textheight]{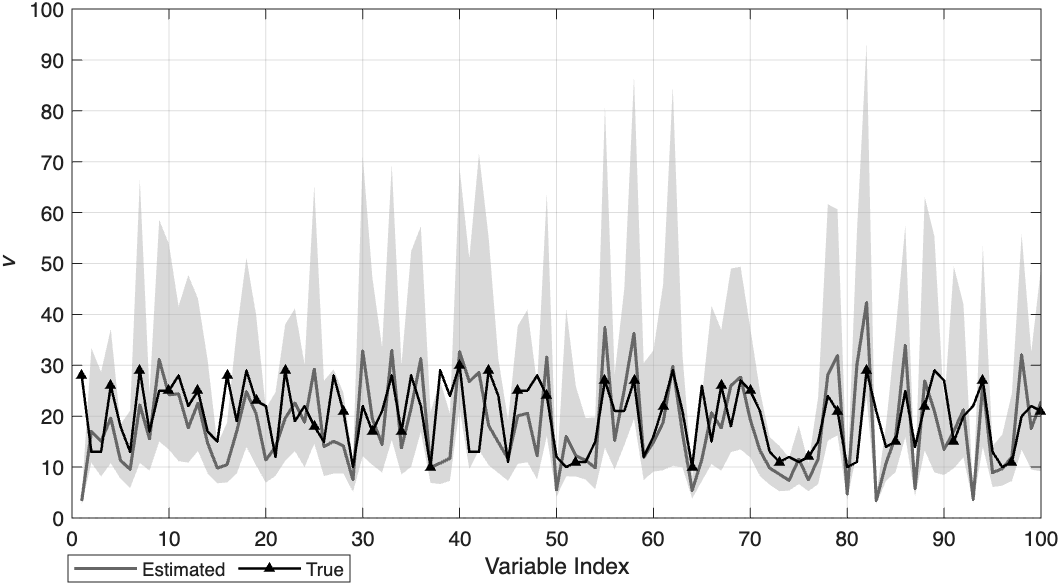}
 \caption*{\footnotesize Notes: The figure displays posterior median estimates (solid line) and true values (black line with triangles) of the degrees-of-freedom parameters $\nu_i$ across variables. Shaded areas denote 90\% credible bands. Lower values of $\nu_i$ correspond to heavier-tailed idiosyncratic innovations.}
\end{figure}

Finally, Figure~\ref{fig:dof_timeseries} reports posterior inference for the degrees-of-freedom parameters governing the idiosyncratic Student-$t$ innovations in line format across variable indices. Posterior medians provide a close approximation to the true degrees of freedom across variables, indicating that the model successfully captures cross-sectional variation in tail thickness. At the same time, the credible bands highlight remaining uncertainty in the estimation of these parameters.

\clearpage

\section{Data}\label{app:data}
This appendix reports the list of U.S. macroeconomic and financial variables used in the forecasting exercise. The dataset is drawn from the FRED-QD database and includes quarterly series covering real activity, labor market conditions, prices, monetary aggregates, interest rates, and financial indicators. All variables are transformed to ensure stationarity following standard conventions as denoted by column "Transformation".

\begin{longtable}{l p{9.5cm} c}
\caption{List of FRED-QD variables used in the forecasting exercise}
\label{tab:data_fredqd}\\
\toprule
FRED mnemonic & Description & Transformation \\
\midrule
\endfirsthead

\multicolumn{3}{l}{\footnotesize Table~\ref{tab:data_fredqd} (continued)}\\
\toprule
FRED mnemonic & Description & Transformation \\
\midrule
\endhead

\midrule
\multicolumn{3}{r}{\footnotesize Continued on next page}\\
\endfoot

\bottomrule
\endlastfoot

GDPC1 & Real Gross Domestic Product & (5) \\
PAYEMS & All Employees: Total Nonfarm & (5) \\
CPIAUCSL & Consumer Price Index: All Items & (6) \\
FEDFUNDS & Effective Federal Funds Rate & (2) \\
PCDGx & Real Personal Consumption Expenditures: Durable Goods & (5) \\
PCESVx & Real Personal Consumption Expenditures: Services & (5) \\
PCNDx & Real Personal Consumption Expenditures: Nondurable Goods & (5) \\
Y033RC1Q027SBEAx & Real Gross Private Domestic Investment: Fixed Investment (Nonresidential) & (5) \\
PNFIx & Real Private Fixed Investment: Nonresidential & (5) \\
PRFIx & Real Private Fixed Investment: Residential & (5) \\
A014RE1Q156NBEA & Shares of Gross Domestic Product: Gross Private Domestic Investment(Change in Private Inventories) & none \\
A823RL1Q225SBEA & Real Government Consumption Expenditures and Gross Investment (Federal) & none \\
FGRECPTx & Real Federal Government Current Receipts & (5) \\
SLCEx & Real Government State and Local Consumption Expenditures & (5) \\
EXPGSC1 & Real Exports of Goods\&Services & (5) \\
IMPGSC1 & Real Imports of Goods\&Services & (5) \\
IPDMAT & Industrial Production: Durable Materials & (5) \\
IPNMAT & Industrial Production: Nondurable Materials & (5) \\
IPDCONGD & Industrial Production: Durable Consumer Goods & (5) \\
IPNCONGD & Industrial Production: Nondurable Consumer Goods & (5) \\
IPBUSEQ & Industrial Production: Business Equipment & (5) \\
IPFINAL & Industrial Production: Final Products (Market Group) & (5) \\
IPCONGD & Industrial Production: Consumer Goods & (5) \\
IPMAT & Industrial Production: Materials & (5) \\
IPB51222S & Industrial Production: Total Index & (5) \\
CUMFNS & Capacity Utilization: Total Industry & none \\
HOUST & Housing Starts: Total New Privately Owned & (5) \\
HOUSTNE & Housing Starts: Northeast & (5) \\
HOUSTMW & Housing Starts: Midwest & (5) \\
HOUSTS & Housing Starts: South & (5) \\
HOUSTW & Housing Starts: West & (5) \\
PERMIT & New Private Housing Units Authorized by Building Permits & (5) \\
PERMITNE & New Private Housing Units Authorized by Building Permits, Northeast & (5) \\
PERMITMW & New Private Housing Units Authorized by Building Permits, Midwest & (5) \\
PERMITS & New Private Housing Units Authorized by Building Permits, South & (5) \\
PERMITW & New Private Housing Units Authorized by Building Permits, West & (5) \\
NAPM & ISM Manufacturing: PMI & none \\
NAPMNOI & ISM Manufacturing: New Orders Index & none \\
NAPMSDI & ISM Manufacturing: Supplier Deliveries Index & none \\
NAPMII & ISM Manufacturing: Inventories Index & none \\
NAPMPRI & ISM Manufacturing: Prices Index & none \\
CES1021000001 & All Employees: Mining and Logging & (5) \\
CES3000000001 & All Employees: Manufacturing & (5) \\
CES4000000001 & All Employees: Trade, Transportation \& Utilities & (5) \\
CES5000000001 & All Employees: Information & (5) \\
CES5500000001 & All Employees: Financial Activities & (5) \\
CES6000000001 & All Employees: Professional \& Business Services & (5) \\
CES6500000001 & All Employees: Education and Health Services & (5) \\
CES7000000001 & All Employees: Leisure \& Hospitality & (5) \\
CES8000000001 & All Employees: Other Services & (5) \\
USGOOD & All Employees: Goods-Producing Industries & (5) \\
USCONS & All Employees: Construction & (5) \\
USFIRE & All Employees: Financial Activities & (5) \\
USPBS & All Employees: Professional \& Business Services & (5) \\
USLAH & All Employees: Leisure \& Hospitality & (5) \\
USSERV & All Employees: Other Services & (5) \\
USMINE & All Employees: Mining and Logging & (5) \\
USTPU & All Employees: Trade, Transportation \& Utilities & (5) \\
USTRADE & All Employees: Retail Trade & (5) \\
USWTRADE & All Employees: Wholesale Trade & (5) \\
CES9091000001 & All Employees: Government--Federal & (5) \\
CES9092000001 & All Employees: Government--State Government & (5) \\
CES9093000001 & All Employees: Government--Local Government & (5) \\
LNS14000012 & Unemployment Rate--16 to 19 Years & (2) \\
LNS14000025 & Unemployment Rate--20 Years and over, Men & (2) \\
LNS14000026 & Unemployment Rate--20 Years and over, Women & (2) \\
UEMPLT5 & Number of Civilians Unemployed--Less Than 5 Weeks & (5) \\
UEMP5TO14 & Number of Civilians Unemployed for 5 to 14 Weeks & (5) \\
UEMP15T26 & Number of Civilians Unemployed for 15 to 26 Weeks & (5) \\
UEMP27OV & Number of Civilians Unemployed for 27 Weeks and Over & (5) \\
LNS12032194 & Employment Level--Part-Time for Economic Reasons, All Industries & (5) \\
AWHMAN & Average Weekly Hours of Production and Nonsupervisory Employees: Manufacturing & none \\
AWHNONAG & Average Weekly Hours Of Production And Nonsupervisory Employees: Total Private & (2) \\
AWOTMAN & Average Weekly Overtime Hours of Production and Nonsupervisory Employees: Manufacturing & none \\
CES0600000007 & Average Weekly Hours of Production and Nonsupervisory Employees: Goods-Producing & none \\
CES0600000008 & Average Hourly Earnings of Production and Nonsupervisory Employees: Goods-Producing & (6) \\
CES0600000003 & Average Hourly Earnings of Production and Nonsupervisory Employees: Total Private & (6) \\
GS10 & 10-Year Treasury Constant Maturity Rate & (2) \\
GS5 & 5-Year Treasury Constant Maturity Rate & (2) \\
GS1 & 1-Year Treasury Constant Maturity Rate & (2) \\
TB3MS & 3-Month Treasury Bill: Secondary Market Rate & (2) \\
BAA & Moody's Seasoned Baa Corporate Bond Yield & (2) \\
AAA & Moody's Seasoned Aaa Corporate Bond Yield & (2) \\
M2SL & M2 Money Stock & (5) \\
M1SL & M1 Money Stock & (5) \\
BUSLOANS & Commercial and Industrial Loans, All Commercial Banks & (5) \\
REALLN & Real Estate Loans, All Commercial Banks & (5) \\
CONSUMER & Consumer Loans, All Commercial Banks & (5) \\
TOTALSL & Total Consumer Credit Outstanding & (5) \\
MORTGAGE30US & 30-Year Fixed Rate Mortgage Average in the United States & (2) \\
PCEPI & Personal Consumption Expenditures: Chain-type Price Index & (6) \\
GDPDEF & Gross Domestic Product: Implicit Price Deflator & (6) \\
PPIACO & Producer Price Index: All Commodities & (6) \\
CPILFESL & Consumer Price Index: All Items Less Food and Energy & (6) \\
DCLORG3Q086SBEA & Personal Consumption Expenditures: Nondurable Goods--Clothing and Footwear & (6) \\
DGOERG3Q086SBEA & Personal Consumption Expenditures: Nondurable Goods--Gasoline and Other Energy Goods & (6) \\
DONGRG3Q086SBEA & Personal Consumption Expenditures: Nondurable Goods--Other Nondurable Goods & (6) \\
DHUTRG3Q086SBEA & Personal Consumption Expenditures: Services--Housing and Utilities & (6) \\
DHLCRG3Q086SBEA & Personal Consumption Expenditures: Services--Healthcare (chain-type price index) & (6) \\
DTRSRG3Q086SBEA & Personal Consumption Expenditures: Transportation Services & (6) \\
DRCARG3Q086SBEA & Personal Consumption Expenditures: Recreation Services & (6) \\
DFSARG3Q086SBEA & Personal Consumption Expenditures: Services--Food Services and Accommodations & (6) \\
DIFSRG3Q086SBEA & Personal Consumption Expenditures: Financial Services and Insurance & (6) \\
DOTSRG3Q086SBEA & Personal Consumption Expenditures: Other Services & (6) \\
WPSFD49502 & Producer Price Index by Commodity for Final Demand: Personal Consumption Goods(Finished Consumer Goods) & (6) \\
WPSFD4111 & Producer Price Index by Commodity for Finished Consumer Foods & (6) \\
PPIIDC & Producer Price Index by Commodity Industrial Commodities & (6) \\
WPSID61 & Producer Price Index by Commodity Intermediate Demand: Processed Goods for Intermediate Demand & (6) \\
WPSID62 & Producer Price Index by Commodity Intermediate Demand: Unprocessed Goods for Intermediate Demand & (6) \\
WPUFD49207 & Producer Price Index by Commodity for Final Demand: Finished Consumer Energy Goods & (6) \\
WPUFD49210 & Producer Price Index by Commodity for Final Demand: Finished Consumer Durable Goods & (6) \\
WPUFD49211 & Producer Price Index by Commodity for Final Demand: Finished Consumer Nondurable Goods & (6) \\
WPUFD49213 & Producer Price Index by Commodity for Final Demand: Finished Consumer Services & (6) \\
OILPRICEx & Crude Oil, West Texas Intermediate (WTI) Spot Price & (5) \\
EXSZUSx & Switzerland/US Foreign Exchange Rate & (5) \\
EXJPUSx & Japan/US Foreign Exchange Rate & (5) \\
EXUSUKx & UK/US Foreign Exchange Rate & (5) \\
EXCAUSx & Canada/US Foreign Exchange Rate & (5) \\
UMCSENTx & University of Michigan: Consumer Sentiment & none \\
SP500 & S\&P500 Common Stock Price Index: Composite & (5) \\

\end{longtable}

\begin{minipage}{\textwidth}
\footnotesize
\textbf{Notes:} Transformation codes are: (2) $\Delta x_t$, (5) $\Delta \log(x_t)$, (6) $\Delta^2 \log(x_t)$. Variables marked ``none'' enter in levels.
\end{minipage}

\section{Forecast evaluation}
\subsection{Benchmark linear dynamic factor model}\label{app:dfm}

This subsection provides a detailed description of the benchmark linear dynamic factor model (DFM) used in the forecasting comparison.
Let $X_{it}$ denote the $i$th observable series at time $t$. The benchmark DFM is
\begin{equation}
X_{it} = B_i^{b}\, f_t^{b} + v_{it}^{b},
\end{equation}
where $f_t^{b}$ is a $J$-dimensional vector of latent factors capturing common macroeconomic fluctuations and $B_i^{b}$ denotes the associated factor loadings. Both the factor loadings and the measurement equation are time invariant. The idiosyncratic component $v_{it}^{b}$ follows a series-specific autoregressive process of order $L_b$,
\begin{equation}
v_{it}^{b}
= \sum_{\ell=1}^{L_b} \rho_{i,\ell}^{b}\, v_{i,t-\ell}^{b}
+ \epsilon_{it}^{b},
\qquad
\epsilon_{it}^{b} \sim \mathcal{N}(0,\omega_{it}^{b}),
\end{equation}
allowing for serial correlation and heteroskedasticity at the series level.

Idiosyncratic volatility is series-specific and evolves according to a stochastic volatility process,
\begin{equation}
\log \omega_{it}^{b}
= \log \omega_{i,t-1}^{b}
+ \xi_{it}^{b},
\qquad
\xi_{it}^{b} \sim \mathcal{N}(0,q_i^{b}),
\end{equation}
so that uncertainty is time-varying but does not contain a common component across series.

The dynamics of the latent factors are governed by a linear VAR($L_b^{f}$),
\begin{equation}
f_t^{b}
= \mu^{b}
+ \sum_{j=1}^{L_b^{f}} \Phi_j^{b}\, f_{t-j}^{b}
+ u_t^{b},
\qquad
u_t^{b} \sim \mathcal{N}(0,\Sigma_f^{b}),
\end{equation}
with constant coefficients and homoskedastic innovations.

\end{document}